\documentclass[12pt]{article}
\usepackage{graphicx}
\usepackage{enumerate} 
\usepackage{amsmath,amssymb}

\renewcommand{\thefootnote}{\fnsymbol{footnote}}

\begin{document}

\title{
\begin{flushright}
{\normalsize
YITP-08-3 \\
KUNS-2123 \\
KEK-CP-206 \vspace{1.2cm}\\}
\end{flushright}
\Large \bf
Determination of $B^* B\pi$ coupling\\
in unquenched lattice QCD
\vspace{0.5cm}}

\author{Hiroshi Ohki$^{a,b}$
\thanks{e-mail: ohki@yukawa.kyoto-u.ac.jp},
Hideo Matsufuru$^{c}$
\thanks{e-mail: hideo.matsufuru@kek.jp},
and Tetsuya Onogi$^{a}$
\thanks{e-mail: onogi@yukawa.kyoto-u.ac.jp}
\vspace{0.4cm}\\
$^a${\it \normalsize
Yukawa Institute for Theoretical Physics, 
Kyoto University, Kyoto 606-8502, Japan} \\
$^b${\it \normalsize 
Department of Physics, Kyoto University, 
Kyoto 606-8501, Japan } \\  
$^c${\it \normalsize
High Energy Accelerator Research Organization (KEK),
Tsukuba 305-0801, Japan } 
\vspace{0.8cm} }

\date{February 12, 2008}

\begin{titlepage}

\maketitle
\thispagestyle{empty}

\begin{abstract}
The $B^* B\pi$ coupling is a fundamental parameter of chiral 
effective Lagrangian with heavy-light mesons and can constrain 
the chiral behavior of $f_B$, $B_B$ and 
the $B\rightarrow \pi l \nu$ form factor in the soft pion limit.
We compute the $B^* B \pi $ coupling 
with the static heavy quark and the $O(a)$-improved Wilson light
quark.
Simulations  are carried out with $n_f=2$ unquenched $12^3\times 24$
lattices at $\beta=1.80$ and $16^3\times 32$
lattices at $\beta=1.95$ generated by CP-PACS collaboration.
To improve the statistical accuracy, we employ the all-to-all
propagator technique and the static quark action with smeared
temporal link variables following the quenched study by
Negishi {\it et al.}.
These methods successfully work also on unquenched lattices,
and determine the $B^*B\pi$ coupling with 1--2\% statistical
accuracy on each lattice spacing.
\end{abstract}

\end{titlepage}

\renewcommand{\thefootnote}{\arabic{footnote}}
\setcounter{footnote}{0}

\section{Introduction}

One of the major subjects in particle physics is to determine
the CKM matrix elements in order to test the standard model and find a 
clue to the physics beyond.
While the precision of the experimental data from $B$ factories having
been improving significantly, 
there are still large uncertainties in the CKM matrix elements due to 
the theoretical errors, which includes those in the lattice
determination of the weak matrix elements for the $B$ mesons.

It is often the case that a symmetry helps to obtain
nonperturbative results in field theories.
For example, the chiral Lagrangian based on the approximate chiral
symmetry can help to understand the quark mass dependence of the light
mesons and also to derive nontrivial relations between different
physical quantities related by the chiral symmetry.
For the $B$ mesons, there is another symmetry called `heavy quark
symmetry' which appears in the limit of infinitely large quark mass.
Based on this symmetry one can construct the heavy meson effective
theory, which gives a systematic description of the heavy-light mesons
including $1/M$ corrections.
Using this effective theory one can understand the light quark mass
dependence of various physical observables of the $B$ meson
weak matrix elements and can also derive nontrivial relations between
different quantities, provided the low energy constants being
determined from some method. 

The heavy meson effective Lagrangian has single low energy constant at
the leading order of the $1/M$ expansion. This constant,
$\hat{g}_b$, is called the 
$B^*B\pi$ coupling. Once the $B^*B\pi$ coupling is determined, the
heavy meson effective 
theory can predict various quantities which are important for CKM 
phenomenology~\cite{Boyd:1994pa}.
For example the light quark mass dependence of the $B$ meson decay
constant and the bag parameter can be determined as  
\begin{eqnarray}
f_{B_d}= F \left( 1 + \frac{3}{4}(1+3\hat{g}_b^2) \frac{m_{\pi}^2}{(4\pi
f_{\pi})^2} \log(m_{\pi}^2/{\Lambda^2})\right) + \mbox{ analytic terms},\\ 
B_{B_d}= B \left( 1 + \frac{3}{4}(1-3\hat{g}_b^2) \frac{m_{\pi}^2}{(4\pi
f_{\pi})^2} \log(m_{\pi}^2/{\Lambda^2})\right) + \mbox{ analytic terms}.
\end{eqnarray}
$F$ and $B$ are the low energy constants associated with these
operators, and correspond to those quantities in the chiral limit
of the light quark.
Also the form factor $f^+(q^2)$ for the semileptonic decay
$B\rightarrow\pi l\nu$ can be expressed
in terms of the $B^*$ meson decay constant $f_{B^*}$ and
$\hat{g}_b$ as
\begin{align}
f^+(q^2)
=-\frac{f_{B^*}}{2f_\pi}
  \left[ \hat{g}_b \left(
      \frac{m_{B^*}}{v \cdot k-\Delta}-\frac{m_{B^*}}{m_B}
  \right) +\frac{f_B}{f_{B^*}} \right],
\end{align}
where $v$ is the velocity of the $B$ meson,
$k$ is the pion momentum, and $\Delta=m_{B^*}-m_{B}$.
Therefore it is quite important to determine the $B^*B\pi$ coupling
very precisely from lattice QCD simulations.
For this purpose, one of the promising approaches
is to use the heavy quark effective theory (HQET) with nonperturbative
accuracy including $1/M$ corrections.
HQET allows systematic treatment  of the $b$
quark in the continuum theory
where $1/M$ corrections can also be systematically
included with nonperturbative accuracy.

Despite its usefulness, it is very difficult in practice
to calculate the matrix
elements for  heavy-light systems with HQET
\cite{deDivitiis:1998kj,Abada:2003un,Becirevic:2005zu}.
This is because in the
heavy-light system the self-energy correction to the static quark
gives a significant
contribution to 
the energy, which results in an exponential growth in time
of the noise to signal ratio of the heavy-light meson correlators.
In fact, recent results of $\hat{g}_\infty$ are  
\begin{align}
 & \hat{g}_\infty = 0.51 \pm 0.03_{\text{stat}} \pm 0.11_{\text{sys}}
 &  \text{for} \quad  n_f = 0~\cite{Abada:2003un}, \\
 & \hat{g}_\infty = 0.51 \pm 0.10_{\text{stat}}  
 &  \text{for} \quad  n_f = 2~\cite{Becirevic:2005zu},
\end{align}
which have about $5\%$ and $15\%$ statistical errors for
quenched and unquenched cases, respectively.
An alternative method which extracts $\hat{g}_\infty$ from
the $B$ quark potential was proposed in Ref.~\cite{Detmold:2007wk},
but such accuracies would not be sufficient to test new physics.  
Therefore significant improvements for statistical precision in HQET 
are needed.
Fortunately the two techniques to reduce the statistical error are
developed recently, which are the 
new HQET action~\cite{DellaMorte:2003mn,Della Morte:2005yc}
with HYP smearing~\cite{Hasenfratz:2001hp}
and  
the all-to-all propagators~\cite{Foley:2005ac} with the low mode
averaging~\cite{DeGrand:2002gm,Giusti:2004yp}.
Negishi {\it et al.}~\cite{Negishi:2006sc} tested applicability of
these methods on a quenched lattice, and found 
that the statistical accuracy is drastically improved as
\begin{equation}
\hat{g}_\infty = 0.517 (16)_{\rm stat.} \ \ \ \text{for}\ \ \  n_f = 0,
\end{equation}
namely at 2\% level, even with a modest number of configurations.

Our final goal is to extend the above strategy to unquenched
simulations and give a precise value of the $B^* B\pi$ coupling 
$\hat{g}_b$ with $2+1$ flavors in the continuum limit. 
In this paper, we study the static $B^* B\pi $ coupling in $n_f=2$ 
unquenched QCD combining two techniques of 
the HYP smeared link and the all-to-all propagators. 
Our purpose is two-fold. The primary purpose is 
to perform the first high precision study of $\hat{g}_{\infty}$ 
in $n_f=2$ unquenched QCD, which serves a
reference point for future studies with better control over the 
systematic errors.
The secondary goal is to understand in what conditions the above
methods apply efficiently.
We observe the dependence of the statistical errors on
the time and the numbers of low-lying eigenmodes, as well as
their behavior against variation of the quark mass
and the lattice spacing.
This will help us to understand in which region of parameters
the method can give good control over the statistical errors,
which will also be useful to precision calculations of other
physics parameters for heavy-light systems.

This paper is organized as follows.
In section \ref{observable}, we describe the method to obtain
$B^*B\pi$ coupling from the $B$ meson matrix element.
Section \ref{Results} explains our simulation details.
In this section we first arrive at our final result for
the $B^*B\pi$ coupling with our best parameter setting.
Then in Section \ref{sec:low_mode_averaging}, the efficiency of
the low mode averaging is examined in detail.
Conclusion is given in Section~\ref{Conclusion}.

\section{Lattice observables}
\label{observable}

The Lagrangian of heavy meson effective theory is given as 
\begin{eqnarray}
L= -\mbox{Tr}\left[\bar{H} iv \cdot D H\right]
+ \hat{g}_b \mbox{Tr}\left[\bar{H}H A_{\mu}\cdot \gamma_{\mu}\gamma_5 
\right] +O(1/M),
\end{eqnarray}
where the low energy constant $\hat{g}_b$ is the $B^*B\pi$ coupling,
$v$ is the four-velocity of the heavy-light meson $B$ or $B^*$,
and $H$, $D_{\mu}$, $A_{\mu}$ are described by the $B$, $B^*$ and $\pi$
fields as 
\begin{eqnarray}
H &=&
  \frac{1}{2}(1+\gamma_{\mu}v_{\mu})(iB\gamma_5+B_{\mu}^*\gamma_{\mu}),
   \hspace{0.7cm}
 \xi=\exp(i\pi/f), \\
 D_{\mu} &=& \partial_{\mu}+\frac{1}{2}
  (\xi^{\dagger}\partial_{\mu}\xi +\xi\partial_{\mu}\xi^{\dagger}),
 \hspace{0.7cm}
 A_{\mu} = \frac{i}{2}
(\xi^{\dagger}\partial_{\mu}\xi -\xi\partial_{\mu}\xi^{\dagger}).
\end{eqnarray}
The $B^* B\pi$ coupling can be obtained from the form factor 
at zero recoil which corresponds to the matrix element 
\begin{align}
\langle B^*(p_{B^*},\lambda)|A_i|B(p_B)\rangle 
|_{\vec{p}_{B^*}=\vec{p}_B=0}
=(m_B+m_{B^*})A_1(q^2=0)\epsilon^{(\lambda)}_i,
\end{align}
where $A_1(q^2=0)$ is the matrix element of the transition from
$B$ to $B^*$ at zero recoil with axial current 
$A_i \equiv \bar{\psi}\gamma_5 \gamma_i \psi$
and $\lambda$ stands for polarization~\cite{deDivitiis:1998kj}.
In the static limit,
\begin{align}
\hat{g}_\infty=A_1(q^2=0)
\end{align}
holds. The matrix element $\langle B^* | A_{\mu} | B \rangle $
at the zero recoil
can be obtained from the ratio of 3-point and 2-point functions, $R(t)$.
\begin{equation}
\frac{ \langle B^*(0)| A_i | B(0) \rangle }{2 m_B} 
= \lim_{t, t_A \rightarrow \infty} R(t,t_A), \\
\end{equation}
where
\begin{equation}
R(t,t_A) = 
\frac{\langle {\cal O}_{B^*}^i(t+t_A) A_i(t_A) {\cal O}_B(0) \rangle }
{\langle {\cal O}_{B}(t+t_A) {\cal O}_B(0) \rangle }
\equiv \frac{C_3(t+t_A)}{C_2(t+t_A)}
\end{equation}
with ${\cal O}_{B}$ and ${\cal O}_{B^*}$ are some operator
having quantum numbers of the $B$ and $B^*$ mesons, respectively.
We apply the smearing technique to enhance the ground state contributions
to the correlators as
\begin{eqnarray}
{\cal O}_B(t,\vec{x}) 
 &=& \sum_{\vec{r}} \phi(\vec{r})\bar{q}(t,\vec{x}+\vec{r})
   \gamma^5 h(t,\vec{x}),\\ 
{\cal O}_{B^*}^i(t,\vec{x})
 &=& \sum_{\vec{r}} \phi(\vec{r})\bar{q}(t,\vec{x}+\vec{r})
  \gamma^i h(t,\vec{x}),
\end{eqnarray}
where $\phi(\vec{x})$ is the smearing function.

The lattice HQET action in the static limit is defined as 
\begin{equation}
S = \sum_x \bar{h}(x)\frac{1+\gamma_0}{2}
\left[
h(x)-U_4^{\dagger}(x-\hat{4})h(x-\hat{4}))
\right],
\end{equation}
where $h(x)$ is the heavy quark field.
The static quark propagator 
is obtained by solving the time evolution equation.
As is well known,
the HQET propagator is very noisy, and it becomes increasingly 
serious as the continuum limit is approached.
In order to reduce the noise,
the Alpha collaboration ~\cite{DellaMorte:2003mn,Della Morte:2005yc}
studied the HQET action in which the link 
variables $U_{\mu}(x)$ are replaced by the smeared links $W_{\mu}(x)$ 
in order to  suppress the power divergence.
They found that the noise of the static heavy-light meson is
significantly suppressed with so-called HYP smearing
\cite{Hasenfratz:2001hp}.

The statistical error is further suppressed by applying
the all-to-all propagator technique developed by the 
TrinLat collaboration~\cite{Foley:2005ac}.
Defining the Hermitian lattice Dirac operator $Q \equiv \gamma_5 D$,
where $D$ is the lattice Dirac operator, 
the quark propagator $S_q(x,y)$ is expressed by the inverse of 
the Hermitian Dirac operator $\bar{Q}=Q^{-1}$ as 
\begin{eqnarray}
S_q(x,y) = \bar{Q}(x,y) \gamma_5.
\end{eqnarray}
We divide the light quark propagator into two parts:  
the low mode part and the high mode part.
The low mode part can be obtained using low
eigenmodes of Hermitian Dirac operator $Q$.
The high mode part can be obtained by the standard random noise
methods with time, color, and spin dilutions.  
With the projection operators into the low and high
mode parts,
\begin{equation}
P_0 = \sum_{i=1}^{N_{\rm ev}} v^{(i)}(x) \otimes v^{(i) \dagger}(y),
\hspace{0.7cm}
P_1 = 1 - P_0,
\end{equation}
respectively, the propagator can be decomposed into two parts as
\begin{equation}
\bar{Q}   = \bar{Q}_0+\bar{Q}P_1, 
\end{equation}
\begin{eqnarray}
\bar{Q}_0(x,y) &=& \sum_{i=1}^{N_{\rm ev}} \frac{1}{\lambda_i}
 v^{(i)}(x) \otimes v^{(i) \dagger}(y),
\label{eq:low_prop}
\\
(\bar{Q}P_1)(x,y)
 &=& \frac{1}{N_r} \sum_r^{N_r}\sum_j
\psi_{[r]}^{(j)}(x)\otimes\eta^{(j)\dagger}_{[r]}(y),
\label{eq:high_prop}
\end{eqnarray}
where $N_r$ is the number of random noise and $j$
is the index for dilution to label the set of time, spin and color sources,
$j=(t_0,\alpha_0,a_0)$.
The low mode part $Q_0$ is constructed from the eigenvectors $v^{(i)}$
with their eigenvalues $\lambda_i$, which are to be obtained
at a preceding stage.
As the random noise vector for the high mode part, 
we adopt the complex $Z_2$ noise.
The random noise vector with dilution is given as
\begin{eqnarray}
\eta^{(j)}_{[r]}(\vec{x},t)^a_\alpha
 =  \eta_{[r]}(\vec{x})^a_\alpha 
\delta_{t, t_0} \delta_{a, a_0} \delta_{\alpha, \alpha_0}.
\end{eqnarray}
$\psi$ is given as 
\begin{eqnarray}
\psi_{[r]}(x) = \sum_y(\bar{Q}P_1)(x,y)\eta_{[r]}(y),
\end{eqnarray}
which is obtained by solving a linear equation
$Q\psi_{[r]}=P_1\eta_{[r]}$.
Further details of the computation methods are given
in Ref.~\cite{Negishi:2006sc}.

Combining these propagators, we can obtain the 2-point functions for
the heavy-light meson which are averaged all over the spacetime.
Similarly, the 3-point functions can be divided into four parts: 
low-low, low-high, high-low and high-high parts.

\section{Results}
\label{Results}

\subsection{Simulation setup}

\begin{table}[tb]
\centering
\begin{tabular}{cccccccc}
\hline\hline
$\beta$ & lattice size & $c_{sw}$ & $a^{-1}$[GeV] & $\kappa$
& $m_\pi$[GeV] & $N_{\rm ev}$ & $N_{conf}$ \\
\hline
1.80 & $12^3\times 24$ & 1.60 & 0.9177(92)
                                  & 0.1409  & 1.06 & 200 & 100\\
     &   &          &             & 0.1430  & 0.90 & 200 & 100\\
     &   &          &             & 0.1445  & 0.75 & 200 & 100\\
     &   &          &             & 0.1464  & 0.49 & 200 & 100\\
\hline
1.95 & $16^3\times 32$ & 1.53 & 1.269(14)
                                 & 0.1375  & 1.13 &   0 & 120\\
      &  &         &             & 0.1390  & 0.92 & 200 & 150\\
      &  &         &             & 0.1400  & 0.76 & 200 & 150\\
      &  &         &             & 0.1410  & 0.54 & 200 & 150\\
\hline\hline
\end{tabular}
\caption{The simulation parameters.
The values of the lattice spacing and the pion mass are from
Ref.~\cite{Ali Khan:2001tx}.
}
\label{tab:parameter}
\end{table}

Numerical simulations are carried out on $12^3\times 24$ lattices 
at $\beta=1.80$ and $16^3\times 32$ lattices at $\beta=1.95$
with two flavors of $O(a)$-improved Wilson quarks 
and Iwasaki gauge action.
We make use of about 100 to 150 gauge
configurations  provided by 
CP-PACS collaboration~\cite{Ali Khan:2001tx} through
JLDG (Japan Lattice DataGrid).
We use the $O(a)$-improved Wilson fermion for the light valence quark
with the masses set equal to the sea quark masses.
We use the static quark action with the HYP smeared links with 
the smearing parameter values 
$(\alpha_1,\alpha_2,\alpha_3)=(0.75,0.6,0.3)$ (HYP1)
\cite{DellaMorte:2003mn,Della Morte:2005yc}.
The $B$ and $B^*$ meson operators are smeared with a function 
$\phi(r)=\exp{(-0.9\,\hat{r})}$, where $\hat{r}$ is the distance between
the heavy quark and the light quark in lattice units.
The configurations are fixed to the Coulomb gauge.

We obtain the low-lying eigenmodes of the Hermitian Dirac operator
using implicitly restarted Lanczos algorithm.
The low mode parts of the correlation functions are
computed with $N_{\rm ev}$ = 200 low-lying eigenmodes, except for
the case of $\kappa=0.1375$ at $\beta=1.95$ which is obtained
with $N_{\rm ev}=0$.
The reason of this choice will be explained
in Sec.~\ref{sec:low_mode_averaging}.
The high mode parts of the correlation functions are computed
with complex $Z_2$ random noise vector with $N_r=1$.
The number of time dilution for 
each configuration are set to $N_{t_0}=24$ at $\beta=1.80$ and 
$N_{t_0}=32$ at $\beta=1.95$, respectively.
This setup is based on the experience from the work 
by Negishi {\it et al}.~\cite{Negishi:2006sc}.

\subsection{Correlation function and effective mass}

\begin{figure}[tb]
\begin{center}
\includegraphics[width=7.5cm,clip]{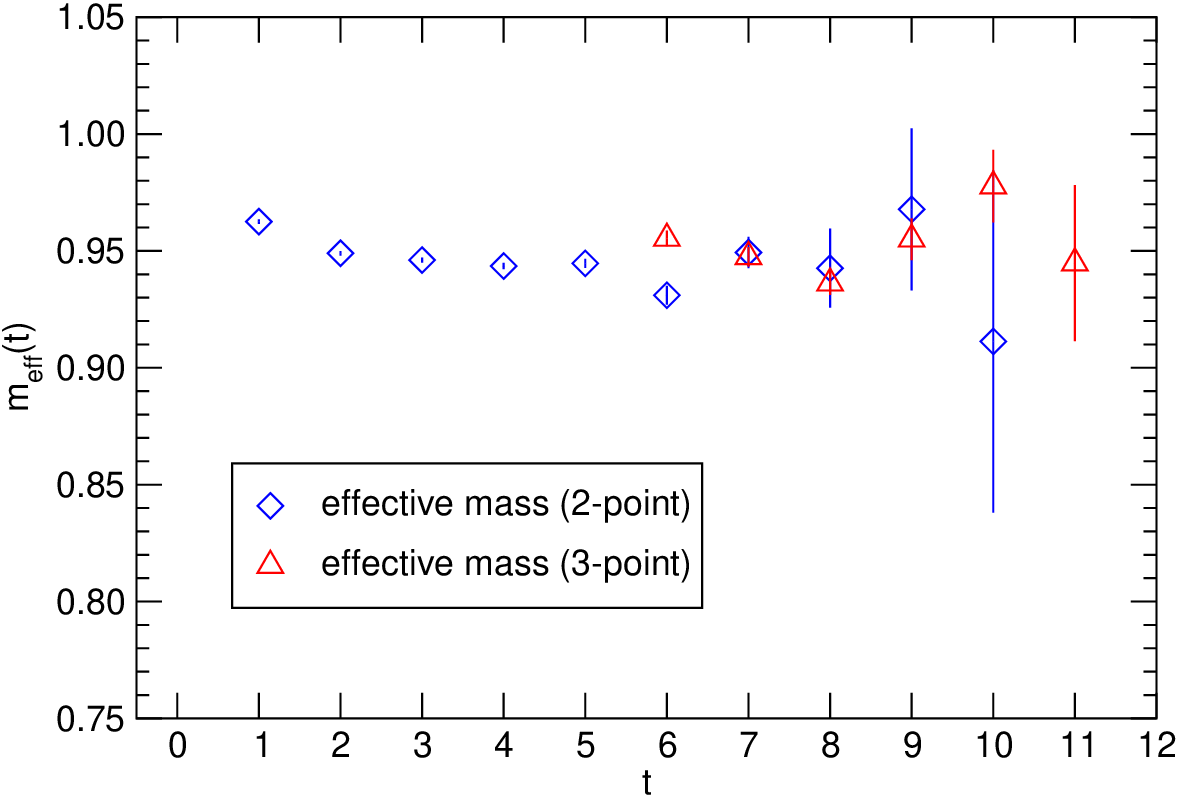}
\includegraphics[width=7.5cm,clip]{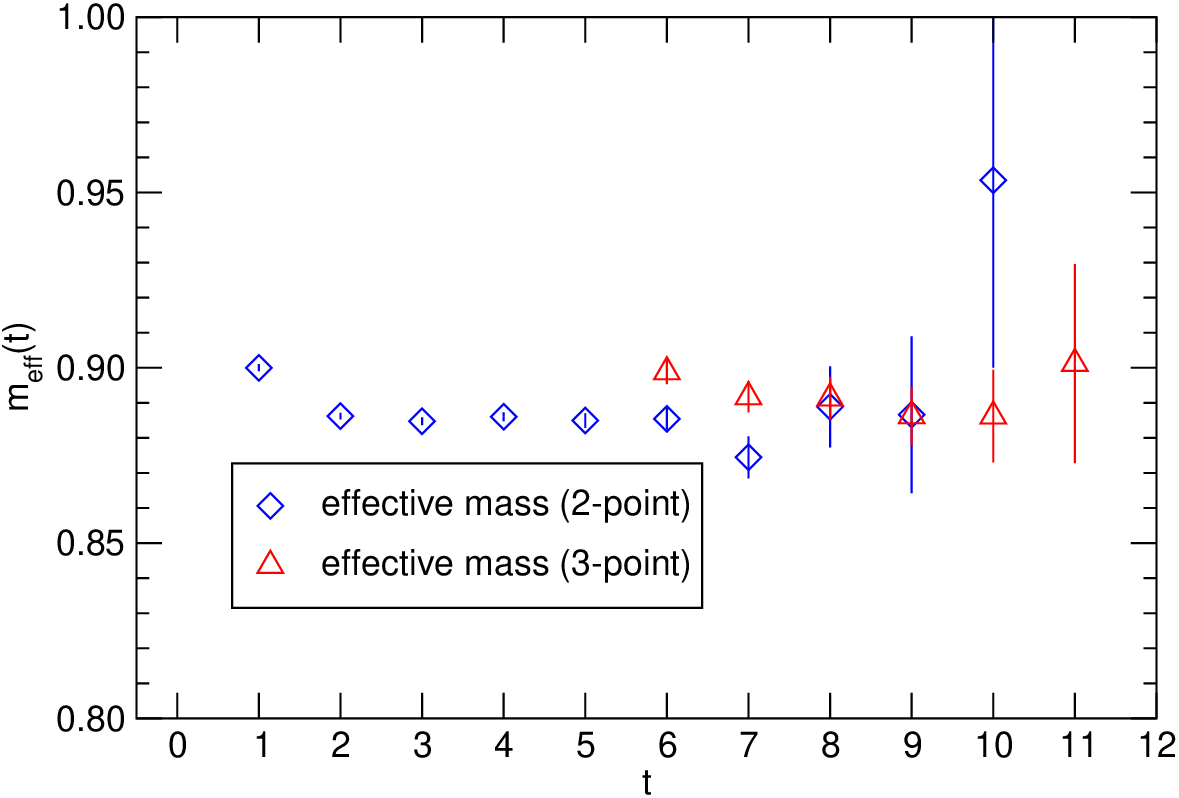}\\
\includegraphics[width=7.5cm,clip]{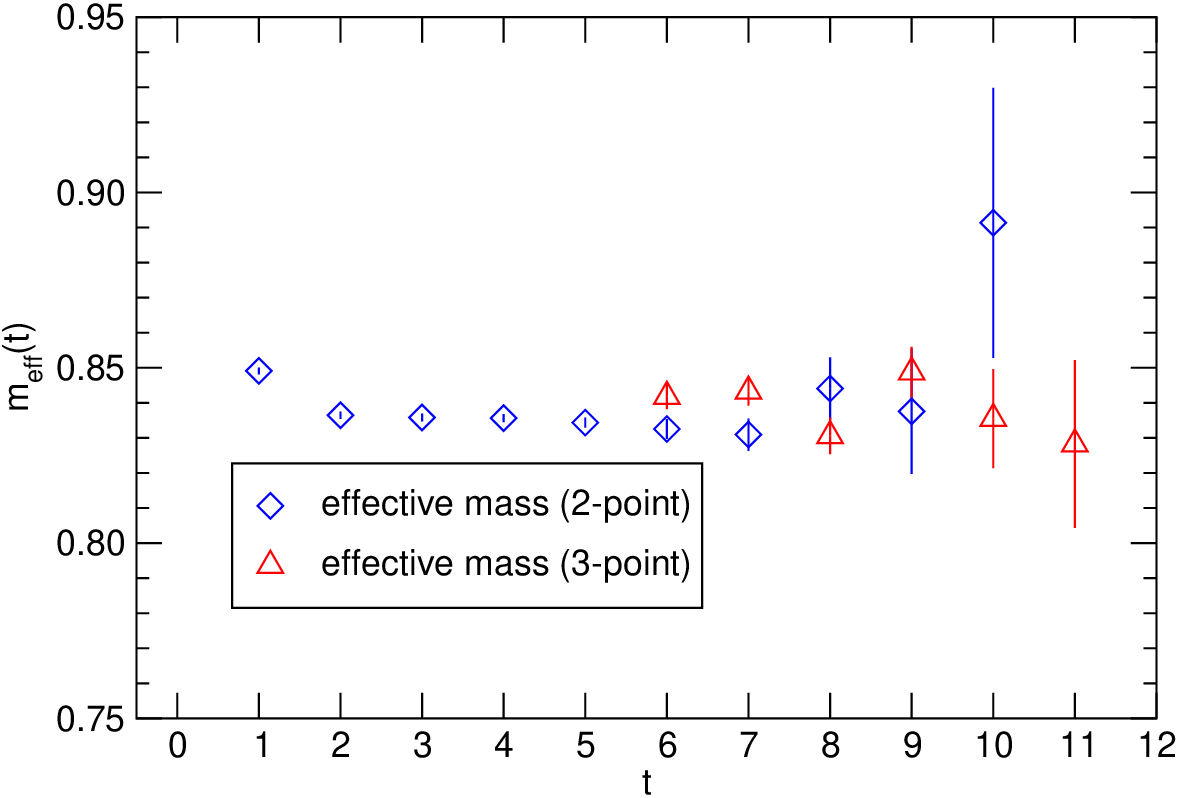}
\includegraphics[width=7.5cm,clip]{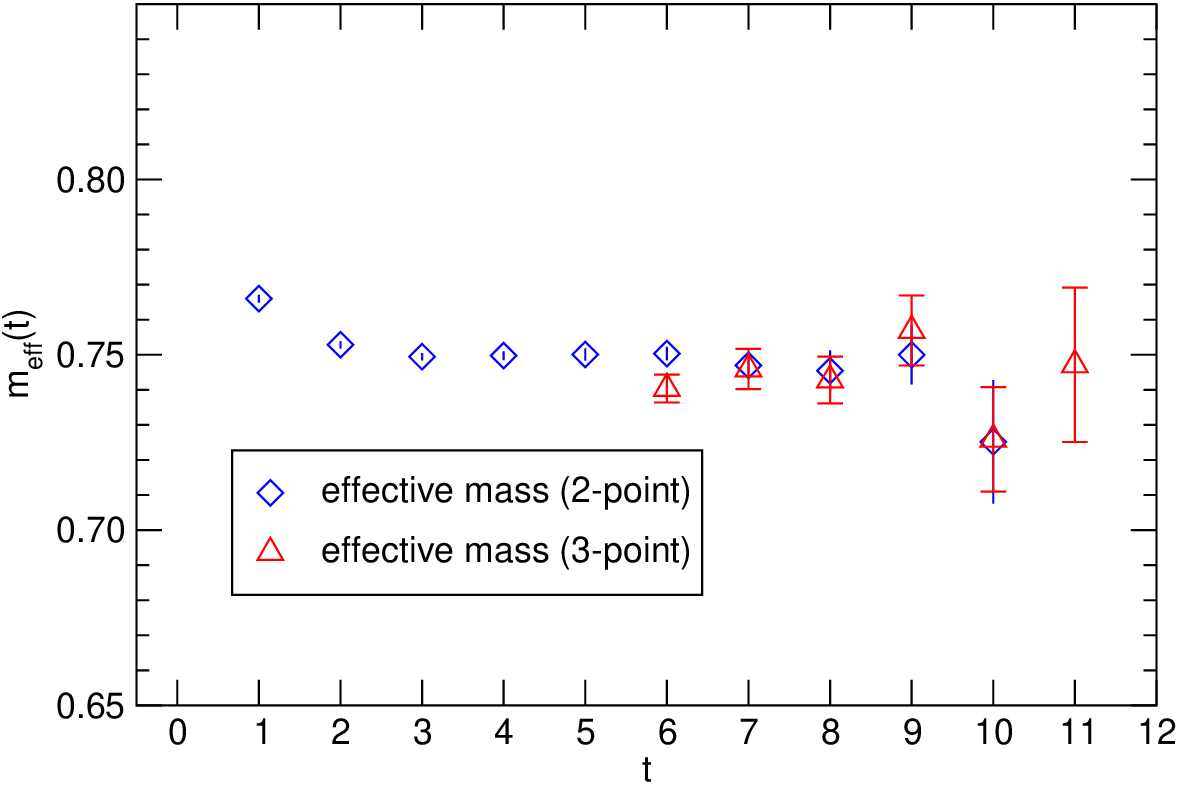}
\caption{
The effective mass plot of the 2-point and 3-point functions at
$\beta=1.80$.
Top left, top right, bottom left, bottom right 
panels correspond to $\kappa=0.1409$, 0.1430, 0.1445, 0.1464,
respectively.}
\label{fig:meff_2pt_b1.80}
\end{center}
\end{figure}

\begin{figure}[tb]
\begin{center}
\includegraphics[width=7.5cm,clip]{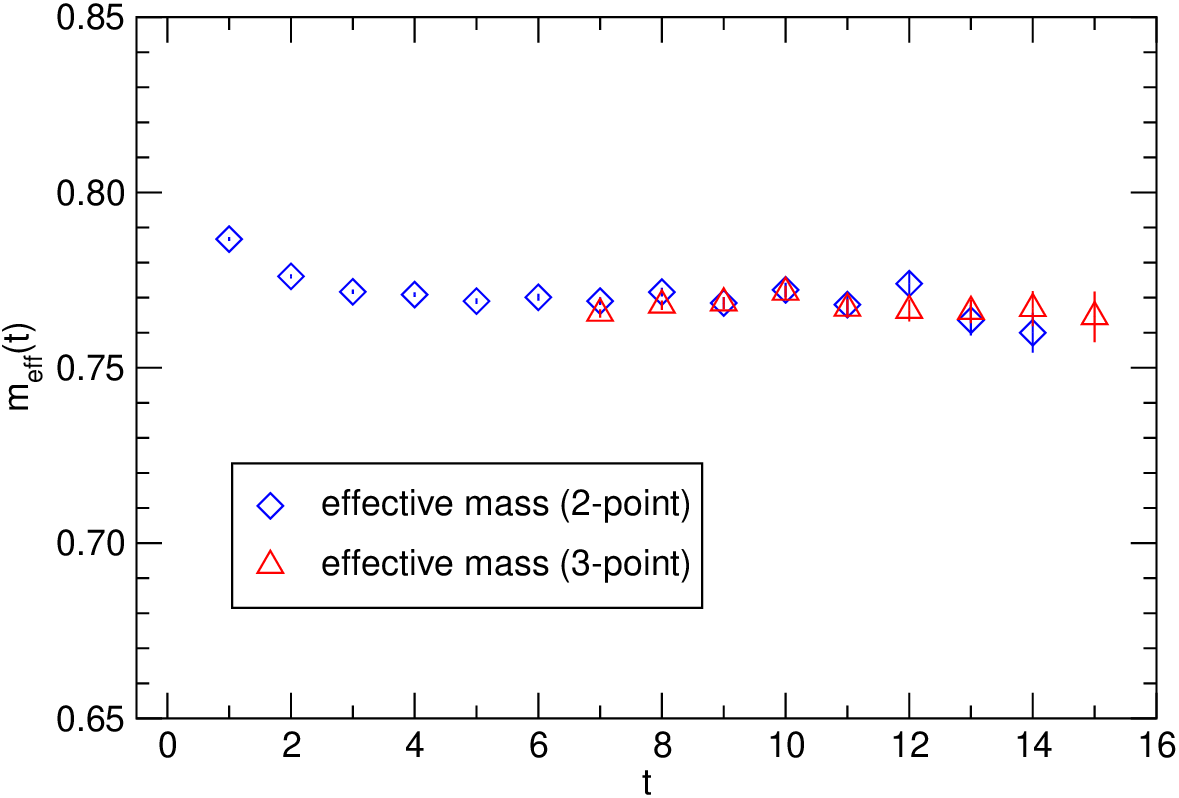}
\includegraphics[width=7.5cm,clip]{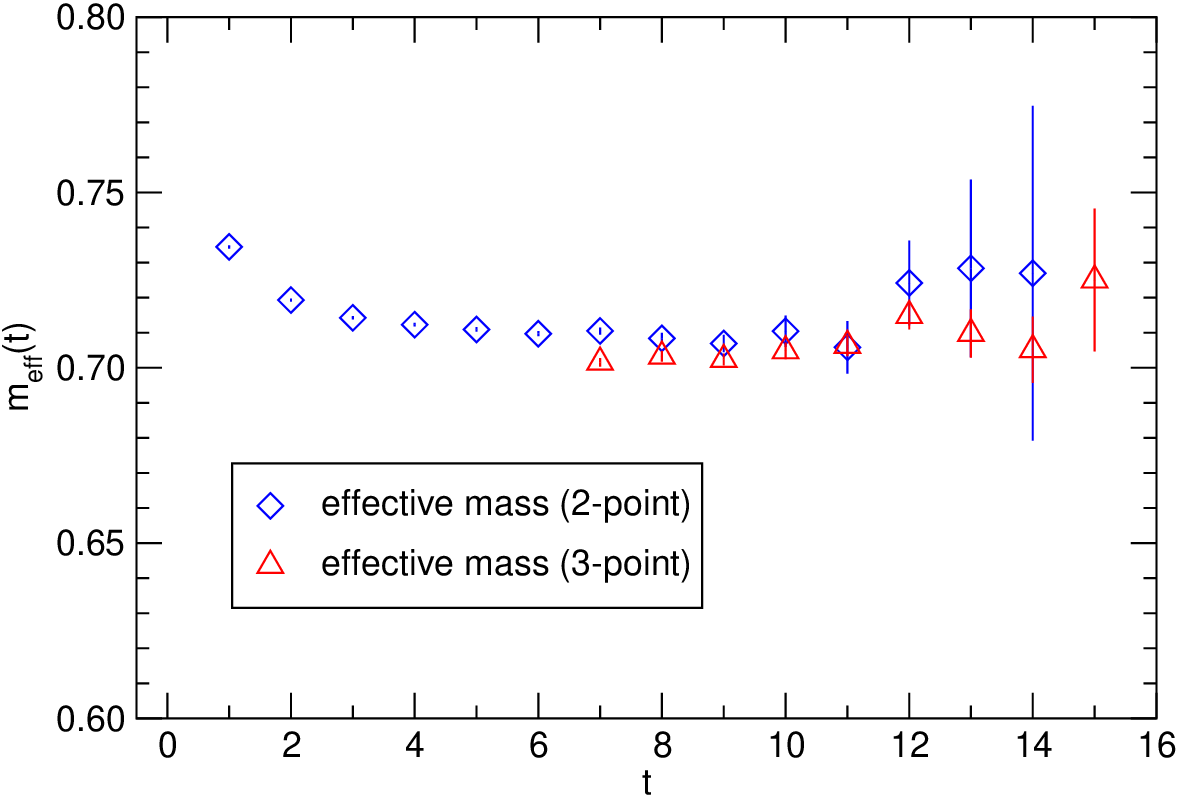} \\
\includegraphics[width=7.5cm,clip]{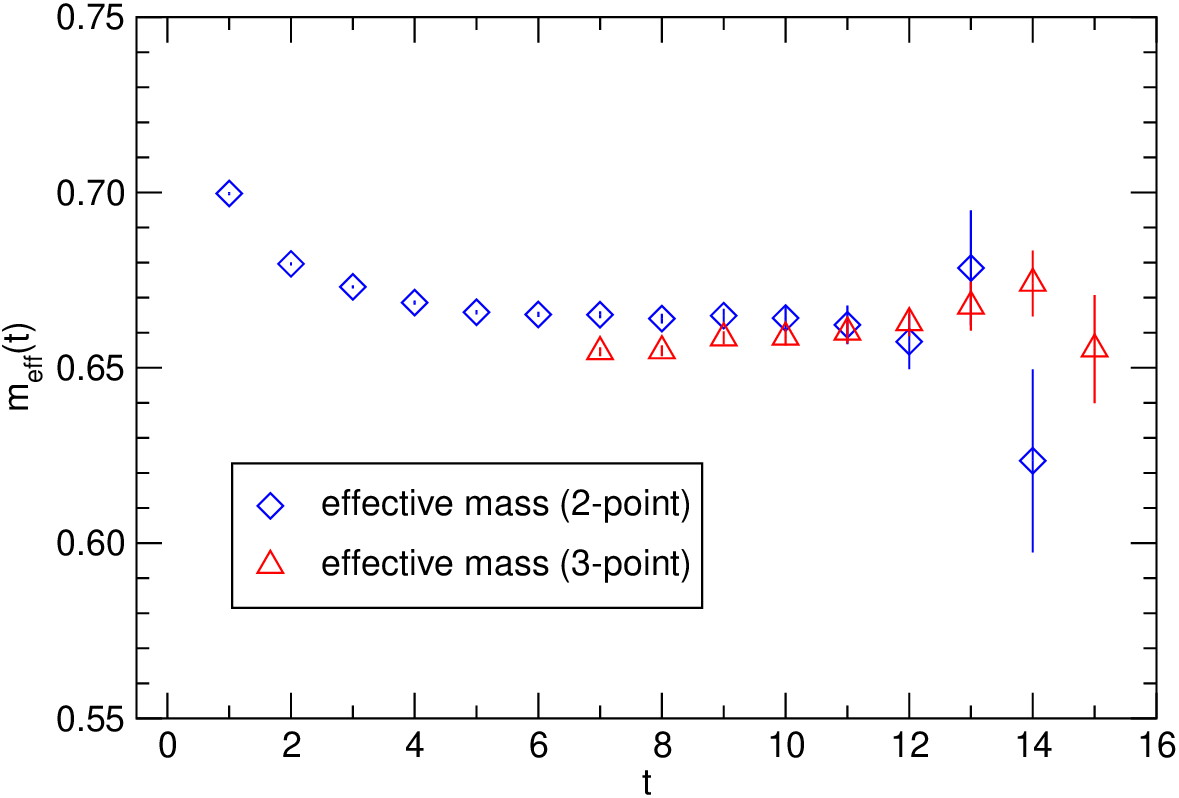}
\includegraphics[width=7.5cm,clip]{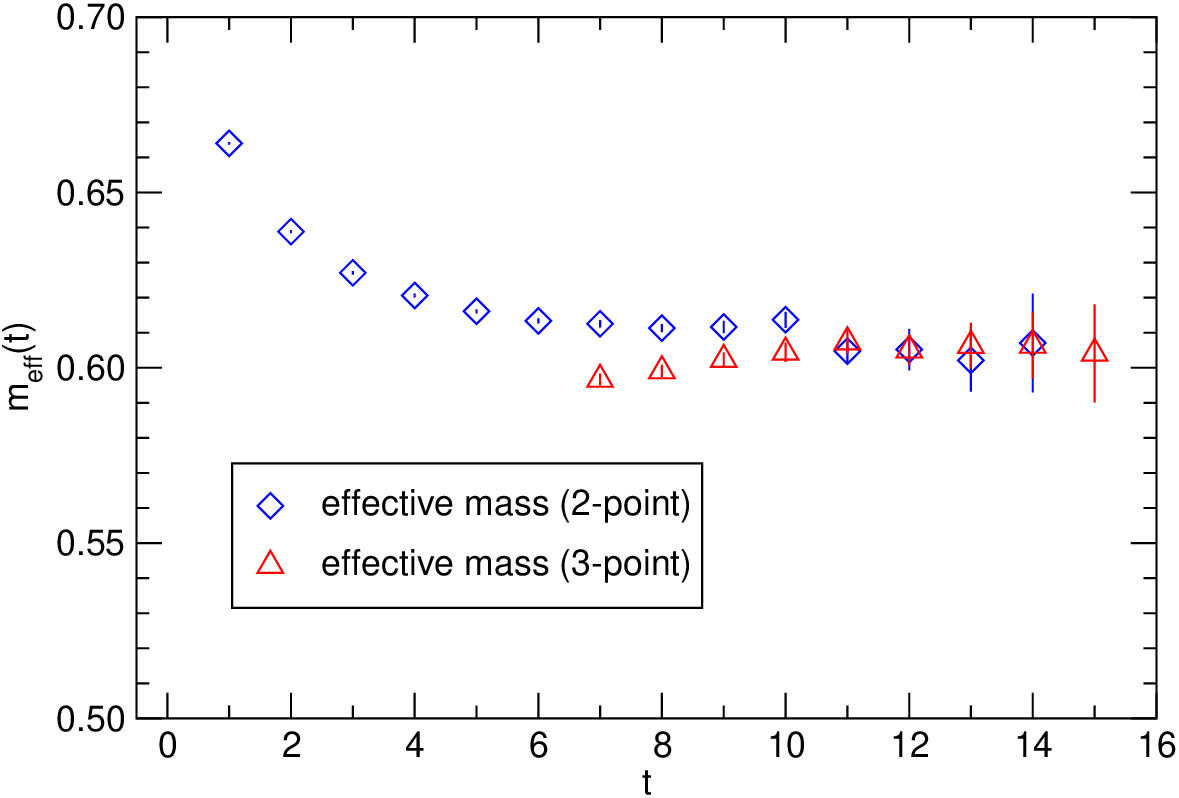}
\caption{
The effective mass plot of the 2-point and 3-point functions at $\beta=1.95$.
Top left, top right, bottom left, bottom right 
panels correspond to $\kappa=0.1375$, 0.1390, 0.1400, 0.1410,
respectively.}
\label{fig:meff_2pt_b1.95}
\end{center}
\end{figure}

\begin{figure}[tb]
\begin{center}
\includegraphics[width=7.5cm,clip]{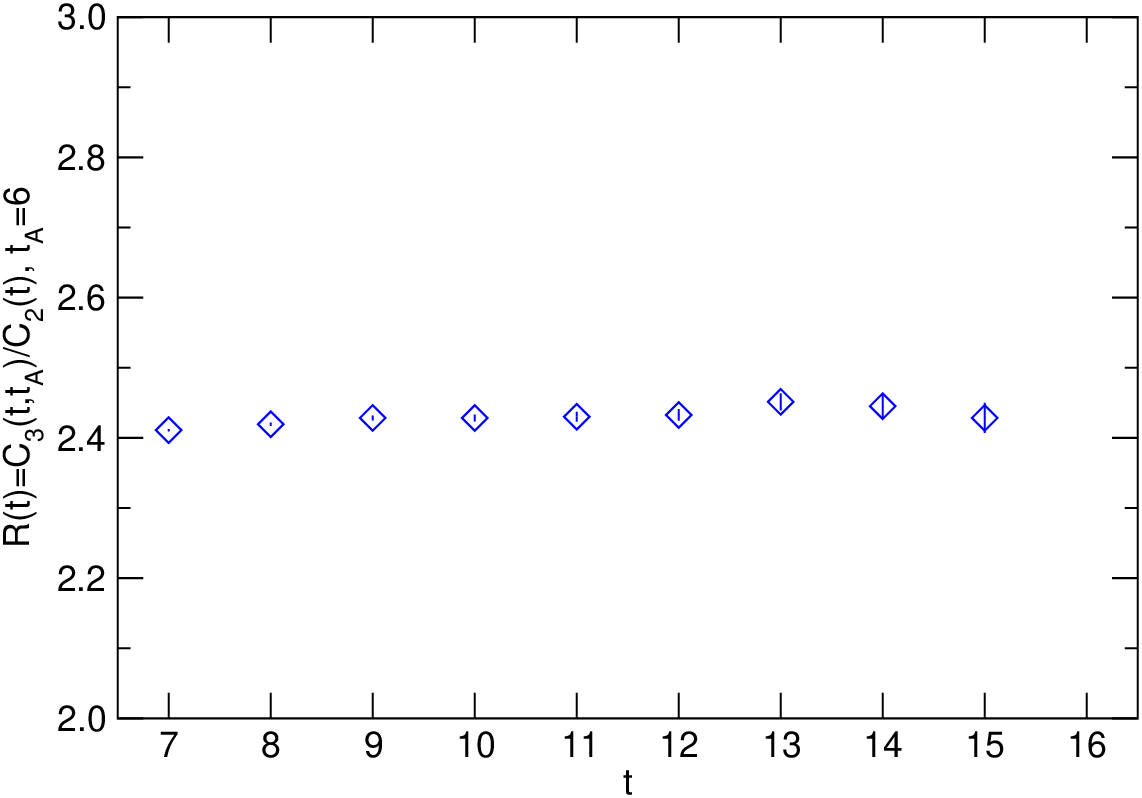}
\includegraphics[width=7.5cm,clip]{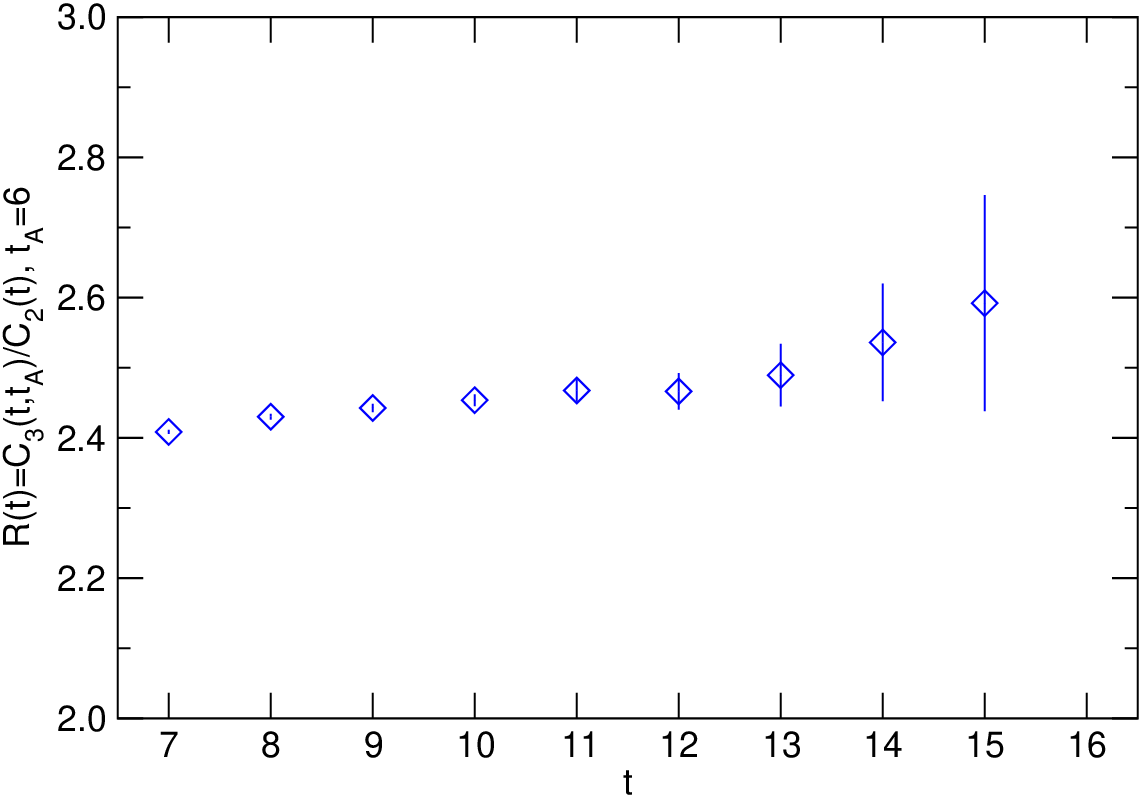}\\
\includegraphics[width=7.5cm,clip]{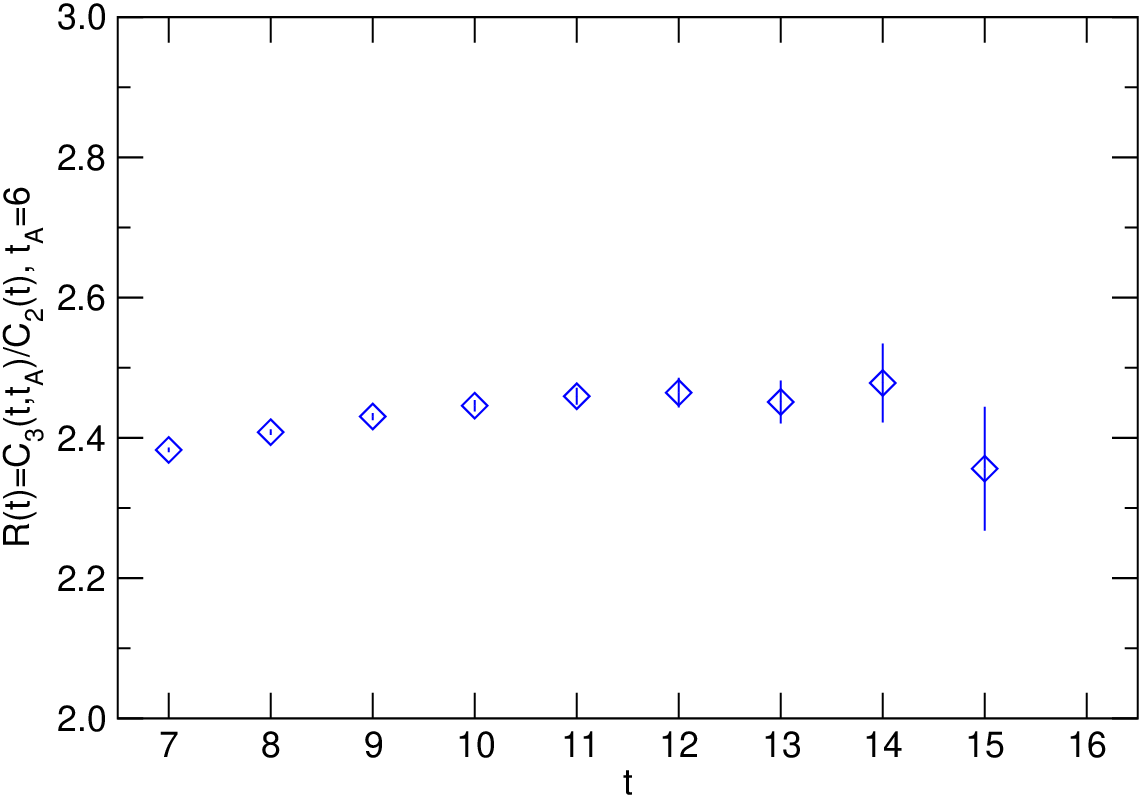}
\includegraphics[width=7.5cm,clip]{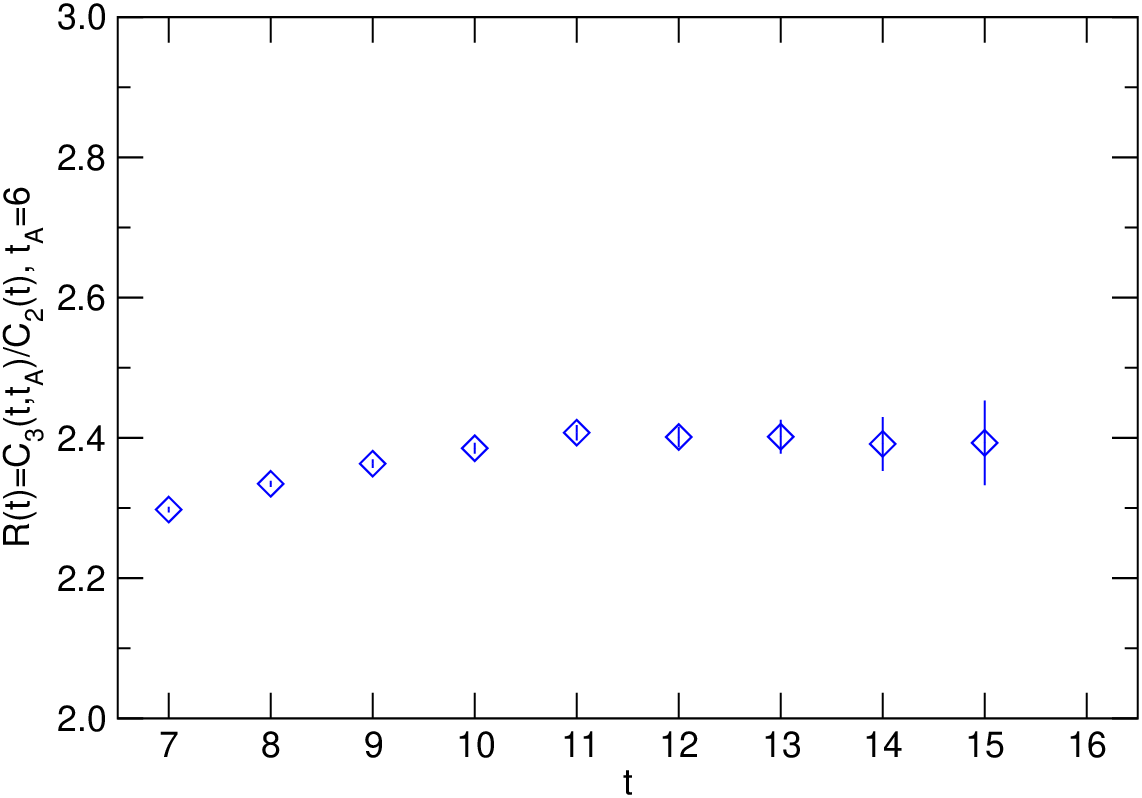}
\vspace{-0.5cm}
\end{center}
\caption{
The ratio of the 3-point and 2-point functions versus
$t$ at $\beta=1.95$.
Correspondence of the panels and the values of $\kappa$
is the same as in Fig.~\ref{fig:meff_2pt_b1.95}.}
\label{fig:ratio_b1.95}
\end{figure}

Figures~\ref{fig:meff_2pt_b1.80}~and~\ref{fig:meff_2pt_b1.95} show the
effective mass plots for 
the 2-point and 3-point functions .
We find that the 2-point functions exhibit nice plateaux at $t \geq 4$
for $\beta=1.80$ and at $t \geq 5$ for 
$\beta=1.95$.
From this result we take $t_A=5$ for $\beta=1.8$ and 
$t_A=6$ for $\beta=1.95$  
as a reasonable 
choice for the time separation between 
the current $A_i$ and the $B$ meson source. 
We also find that the effective masses of the 3-point
functions give consistent values  with those of the 2-point functions.
We fit the 2-point and 3-point functions to exponential functions
with single exponent as 
\begin{equation}
\begin{split} 
 C_2(t)=Z_2 \exp{(-E_{stat}t)}, \hspace{0.7cm}
 C_3(t)=Z_3 \exp{(-E_{stat}t)},
\end{split}
\end{equation}
where $Z_2$ and $Z_3$ are constant parameters and $E_{stat}$
corresponds to the energy of the heavy-light meson.
The fit ranges are chosen appropriately by observing the effective
mass plots as listed in Table~\ref{tab:mass_gphys}.
The bare $B^* B\pi$ coupling can be
obtained by the ratio of the fit parameters as 
$\hat{g}_\infty^{\rm bare}=Z_3/Z_2$ .
Alternatively, the $B^*B\pi$ coupling is also extracted from
the ratio of the 3-point and 2-point functions, $C_3(t)/C_2(t)$,
as shown in Figures~\ref{fig:ratio_b1.95}.
We find that the fit of the ratio $C_3(t)/C_2(t)$ to a constant
value gives consistent value with the value of $Z_3/Z_2$. 
Since the statistical accuracy is better, we employ
$Z_3/Z_2$ to determine $\hat{g}_\infty$ in the following analyses. 
The results are summarized in Table~\ref{tab:mass_gphys}.

\begin{table}[tb]
\centering
\begin{tabular}{ccccccc}
\hline\hline
$\beta$ & $\kappa$ &
$(t^{\rm 2pt}_{\rm min},t^{\rm 2pt}_{\rm max})$ & 
$(t^{\rm 3pt}_{\rm min},t^{\rm 3pt}_{\rm max})$ &
$a E_{stat}$ & $Z_3/Z_2$ & $\hat{g}_\infty$ \\
\hline
1.80 & 0.1409 & (5,10) & (8,10) & 0.9412(19) & 2.252(21)  & 0.612(5) \\
     & 0.1430 & (5,10) & (8,10) & 0.8839(21) & 2.294(23)  & 0.598(5) \\
     & 0.1445 & (5,10) & (8,10) & 0.8343(16) & 2.342(13)  & 0.591(4) \\
     & 0.1464 & (5,10) & (8,10) & 0.7488(17) & 2.381(27)  & 0.578(5) \\
\hline
1.95 & 0.1375 & (8,14) & (11,14)& 0.7669(27) & 2.435(~8)& 0.618(8) \\
     & 0.1390 & (8,14) & (11,14)& 0.7093(18) & 2.471(16)& 0.615(5) \\
     & 0.1400 & (8,14) & (11,14)& 0.6638(15) & 2.461(14)& 0.599(4) \\
     & 0.1410 & (8,14) & (11,14)& 0.6098(14) & 2.400(13)& 0.571(4) \\
\hline\hline
\end{tabular}
\caption{The numerical results of the heavy-light meson energy $aE$,
the ratio of the 3-point and 2-point functions $Z_3/Z_2$,
and $\hat{g}_\infty$.
$(t^{\rm 2pt}_{\rm min},t^{\rm 2pt}_{\rm max})$ and 
$(t^{\rm 3pt}_{\rm min},t^{\rm 3pt}_{\rm max})$ are the fit ranges
for the 2-point and 3-point correlators, respectively.
For the values of $\hat{g}_\infty$, only the statistical errors
are quoted.}
\label{tab:mass_gphys}
\end{table}

\subsection{Physical value of the $B^* B\pi$ coupling 
and chiral extrapolation}

The physical value of the $B^* B\pi$ coupling is obtained 
by multiplying the bare value by the renormalization constant.
We use the one-loop
result of the renormalization factor for the axial vector 
current
\begin{gather}
A_i  = 2\kappa u_0 Z_A 
\left(
1+b_A\frac{m}{u_0} 
\right)
A_i^{lat}, \\
u_0 = 
\left(
1-\frac{0.8412}{\beta}
\right)^{\frac{1}{4}}, \ \ \ \ 
 b_A = 1+ 0.0378 g_{\overline{\rm MS}}^2(\mu), \nonumber
\end{gather}
where the gauge coupling 
$g_{\overline{\rm MS}}^2(\mu)=3.155$, 2.816 and 
$Z_A = 0.932$, 0.939 for $\beta=1.80$ and 1.95,
respectively, as given in Ref.~\cite{Ali Khan:2001tx}.
We arrive at the results of $\hat{g}_\infty$ 
for our $\kappa$ values in Table~\ref{tab:mass_gphys}.

\begin{figure}[tb]
\vspace*{0.5cm}
\begin{center}
\includegraphics[width=7.4cm,clip]{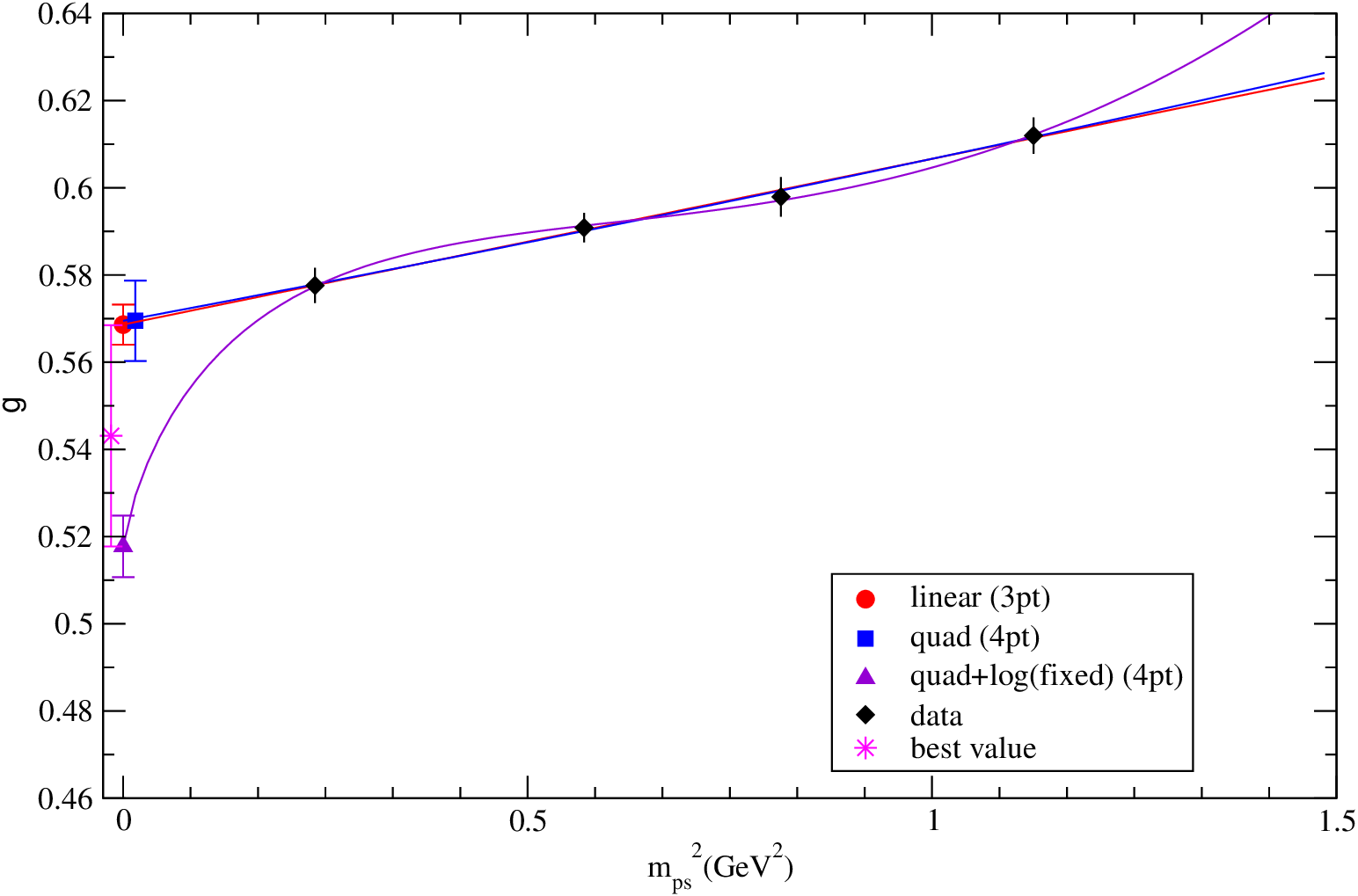}
\includegraphics[width=7.8cm,clip]{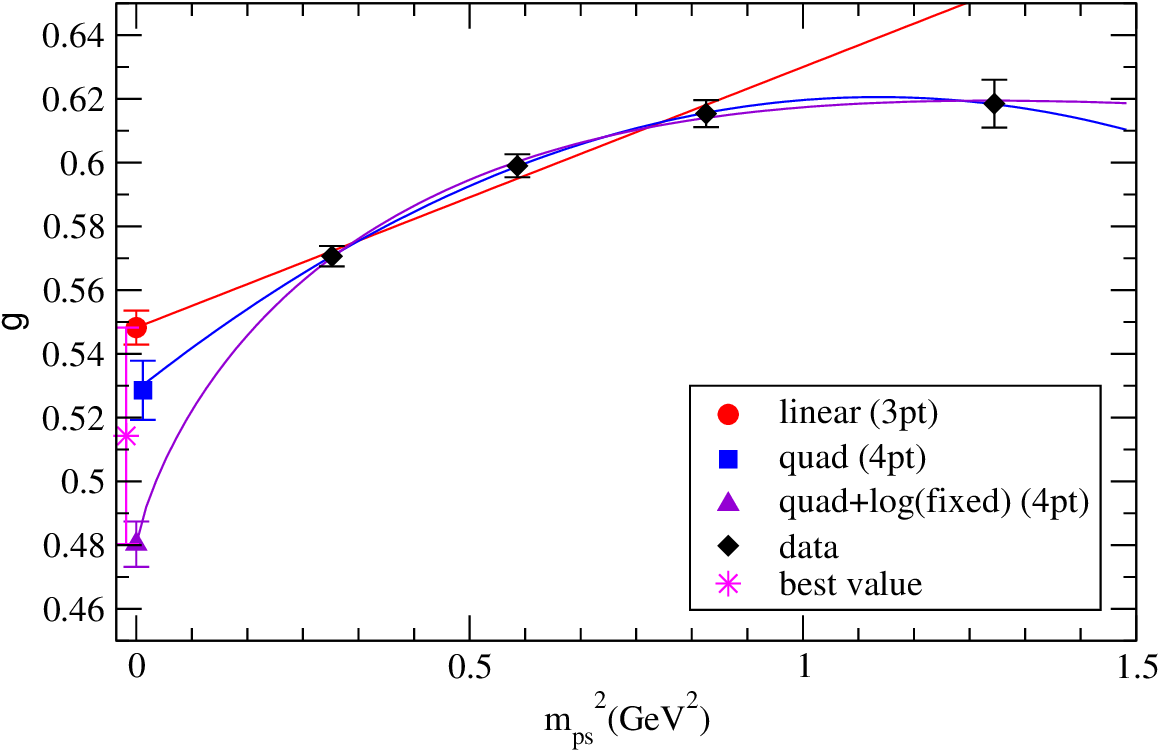}
\caption{
The chiral extrapolation of the physical $B^*B\pi$ coupling 
at $\beta=1.80$ (left panel) and $\beta=1.95$ (right).
}\label{fig:chextr12_16}
\end{center}
\end{figure}

We take the chiral extrapolation of the $B^* B\pi$ 
coupling employing the following fit functions:
\begin{eqnarray}
{\rm (a)}& &  g^{(a)}(m_\pi^2)  = g(0)+A_1 m_\pi^2, \nonumber \\
{\rm (b)}& &  g^{(b)}(m_\pi^2)  = g(0)+A_1 m_\pi^2 +A_2 (m_\pi^2)^2,
 \nonumber \\
{\rm (c)}& &  g^{(c)}(m_\pi^2)  = 
g(0)\left[
 1-g(0)^2\frac{1}{8\pi^2}\frac{m_\pi^2}{f_\pi^2}
\log{(m_\pi^2)}
\right]
+A_1 m_\pi^2 +A_2 (m_\pi^2)^2, \nonumber 
\end{eqnarray}
corresponding respectively to (a) the linear extrapolation,
(b) the quadratic extrapolation, and (c) the quadratic plus chiral
log extrapolation where the log coefficient is determined from ChPT
\cite{Cheng:1993kp,Kamenik:2007md,Fajfer:2006hi}.
We use three lightest data points for the fit (a), while all the four
points for (b) and (c).
We obtain physical values of the $B^* B \pi$ coupling 
in the chiral limit as  
$\hat{g}_\infty = 0.57(1)$, 0.57(2), 0.52(1) at $\beta=1.80$
and 
$\hat{g}_\infty = 0.548(6)$, 0.529(10), 0.480(8) at $\beta=1.95$
from the fit (a), (b), (c), respectively.
Figure~\ref{fig:chextr12_16} shows these chiral extrapolations.
We take the average of the results from the linear fit and
the quadratic plus chiral log fit as our best value and take
half the difference as the systematic error from the chiral
extrapolation:
\begin{eqnarray}
\hat{g}(m_\pi =0)
&=&  0.543(5)_{\text{stat}}(26)_{\text{chiral}}
 \hspace{0.7cm} \text{at}\ \ \beta=1.80,
\\
&=&  0.516(5)_{\text{stat}}(31)_{\text{chiral}}
 \hspace{0.7cm} \text{at}\ \ \beta=1.95.
\end{eqnarray}

Since we have only two lattice spacings,
naive continuum extrapolation would not give a reliable result.
However, the results at these two lattice spacings are consistent 
within quoted errors.%
\footnote{ We do not observe a large scaling violation 
for $\hat{g}_{\infty}$ as opposed to the case of $f_{\pi}$ by CP-PACS
collaboration. A possible explanation is that the large scaling
violation  for $f_{\pi}$ might come from the perturbative error of
$c_A$, which is the $O(a)$-improvement coefficient of the light-light
axial vector current.
If this is the case, $f_{\pi}$ receives significant systematic
error from the $O(a)$-improvement term for the temporal axial vector
current $a c_A \partial_4 P$ due to the chiral enhancement, whereas 
$\hat{g}_{\infty}$ does not receive such a systematic error since the
corresponding term $a c_A \partial_i P$ drops out from the matrix
element by the sum over the space. }
Therefore, we take the result at $\beta=1.95$ 
as our best estimate for the physical value of $\hat{g}_\infty$,
and estimate the discretization error of $O((a\Lambda)^2)$
by order counting with $\Lambda \sim 0.3$ GeV.
Including the perturbative error of $O(\alpha^2)$ also by
order counting, our results for $\hat{g}_\infty$ is 
\begin{equation}
\hat{g}_\infty^{n_f=2} 
=  0.516(5)_{\text{stat}}(31)_{\text{chiral}}
        (28)_{\text{pert}}(28)_{\text{disc}} .
\label{eq:final_result}
\end{equation}

In our study, each of the chiral extrapolation error,
the perturbative error, and the discretization error is about 6\% level. 
The perturbative error can be removed by employing the nonperturbative 
renormalization such as the RI-MOM scheme, which is successfully applied
to the light-light axial vector current.
The discretization error could be reduced  by computing
on finer lattices.
For example, an order counting estimate suggests that the discretization
error would be reduced to about 2\% on the configurations of CP-PACS
at $\beta=2.10$.
In contrast, it is not straightforward to control the chiral
extrapolation error.
It is definitely necessary to use recent unquenched configurations with
smallest pion mass $m_{\pi}\sim 0.3$ GeV.
More predominant approach is to employ a fermion formulation possessing
the chiral symmetry, such as the overlap fermions, which makes
the extrapolation theoretically more transparent.

\section{Applicability of the low mode averaging}
\label{sec:low_mode_averaging}

In this section, we examine under what condition the all-to-all
propagator technique, in particular the low mode averaging,
is efficient to reduce the statistical error.
This would give us a guide to extend our computation to
unquenched simulations with smaller quark masses and finer
lattices.
We mainly investigate the case of $\beta=1.80$ in the following.

\subsection{Observation of the noise to signal ratio}

\begin{figure}[tb]
\begin{center}
\includegraphics[width=8cm,clip]{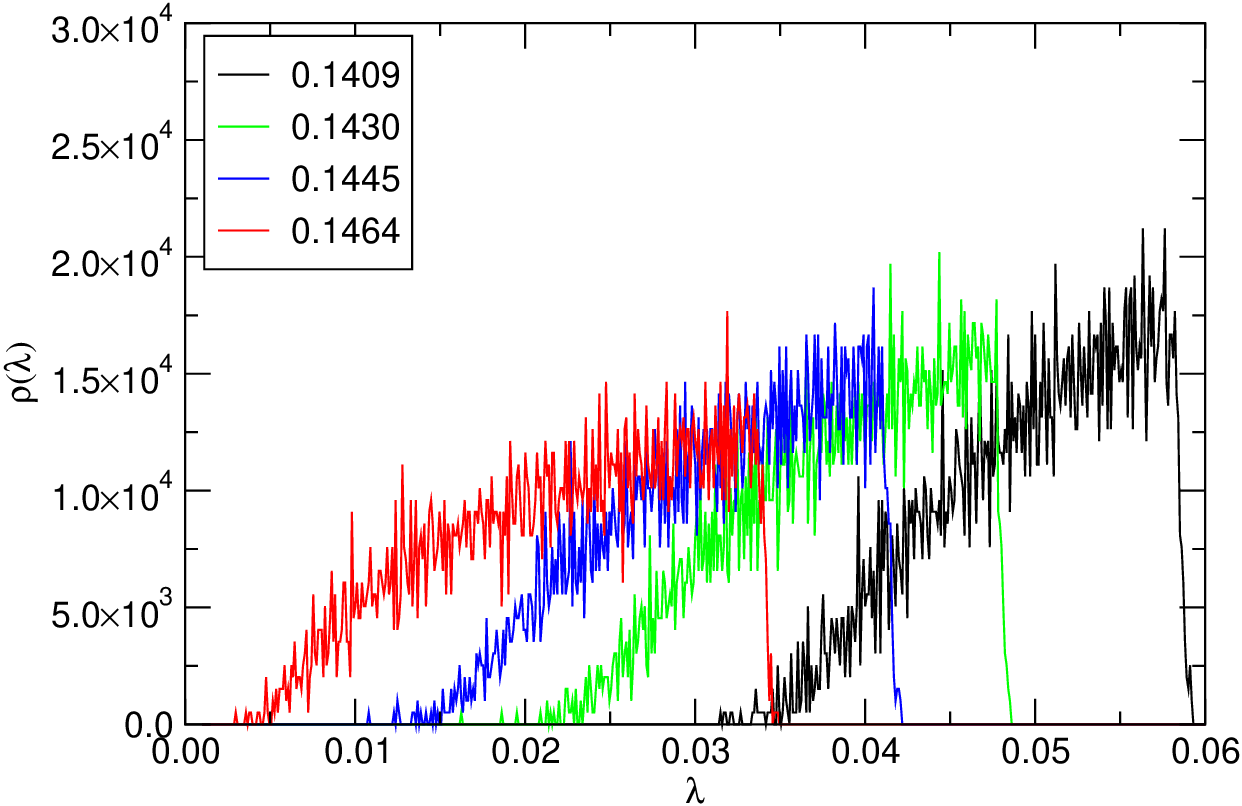}
\vspace{-0.5cm}
\caption{
The low-lying eigenmode distribution for various $\kappa$
at $\beta=1.80$ with 40 configurations. }
\label{fig:hist}
\end{center}
\end{figure}

\begin{figure}[tb]
\includegraphics[width=7.5cm,clip]{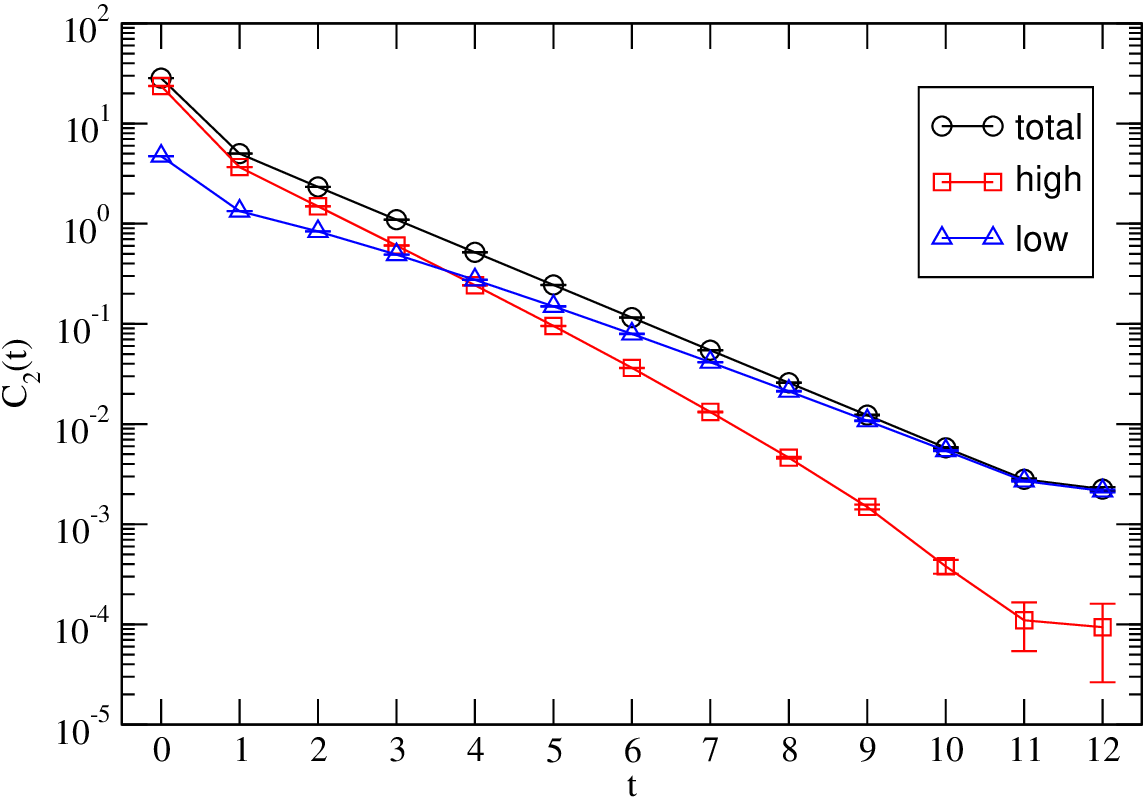}
\includegraphics[width=7.5cm,clip]{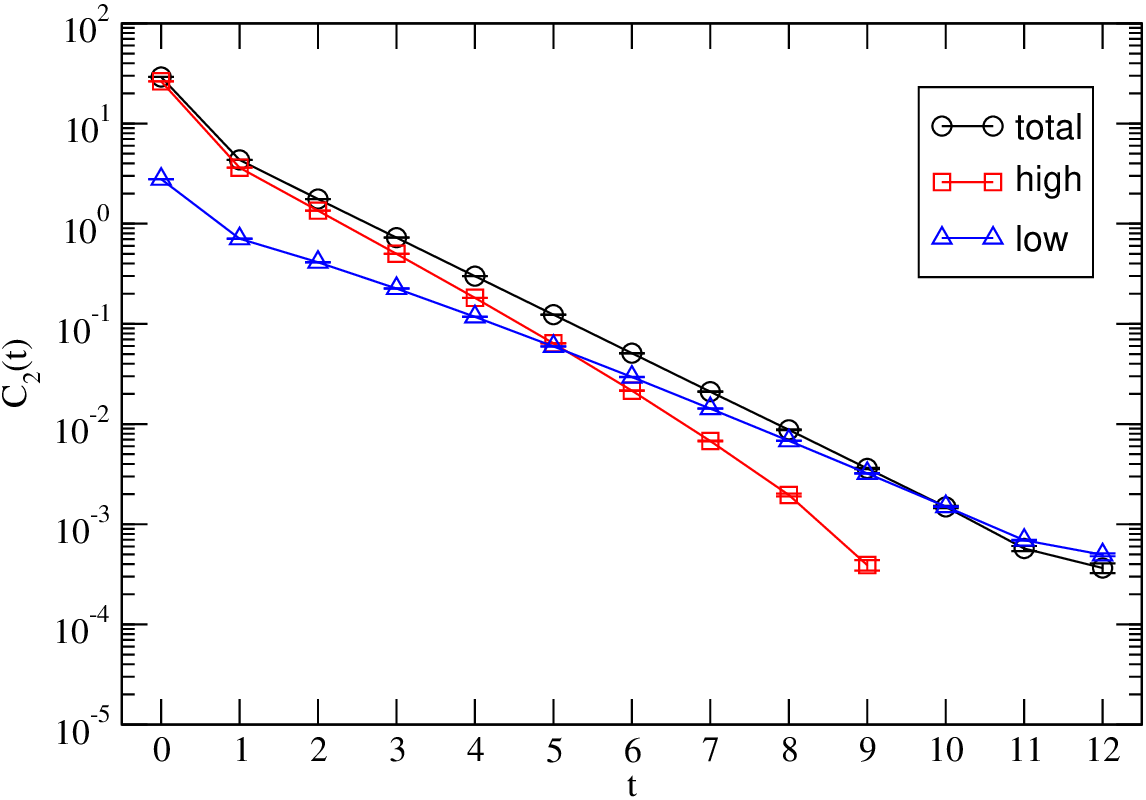}
\caption{
The low and high mode contributions to the 2-point correlators
versus $t$ at $\beta=1.80$.
The left and right panels show the results at $\kappa=0.1464$ and 0.1430,
respectively.}
\label{fig:2pt} 
\end{figure}

Figure~\ref{fig:hist} shows the distribution of about 250
lowest eigenmodes for each $\kappa$.
Since the contribution of each mode $v^{(i)}$ to the correlator
is multiplied by $1/\lambda_i$, one naively expects that
with a fixed number of modes the low mode averaging should be
particularly effective for the smallest quark mass.
This is indeed true for the 2-point and 3-point
heavy-light meson correlators.
Figure~\ref{fig:2pt} represents the low and high mode
contributions to the 2-point correlators.
For the smaller quark mass, the low mode part is indeed more dominant
the 2-point correlator in the earlier $t$.

\begin{figure}[tb]
\includegraphics[width=7.5cm,clip]{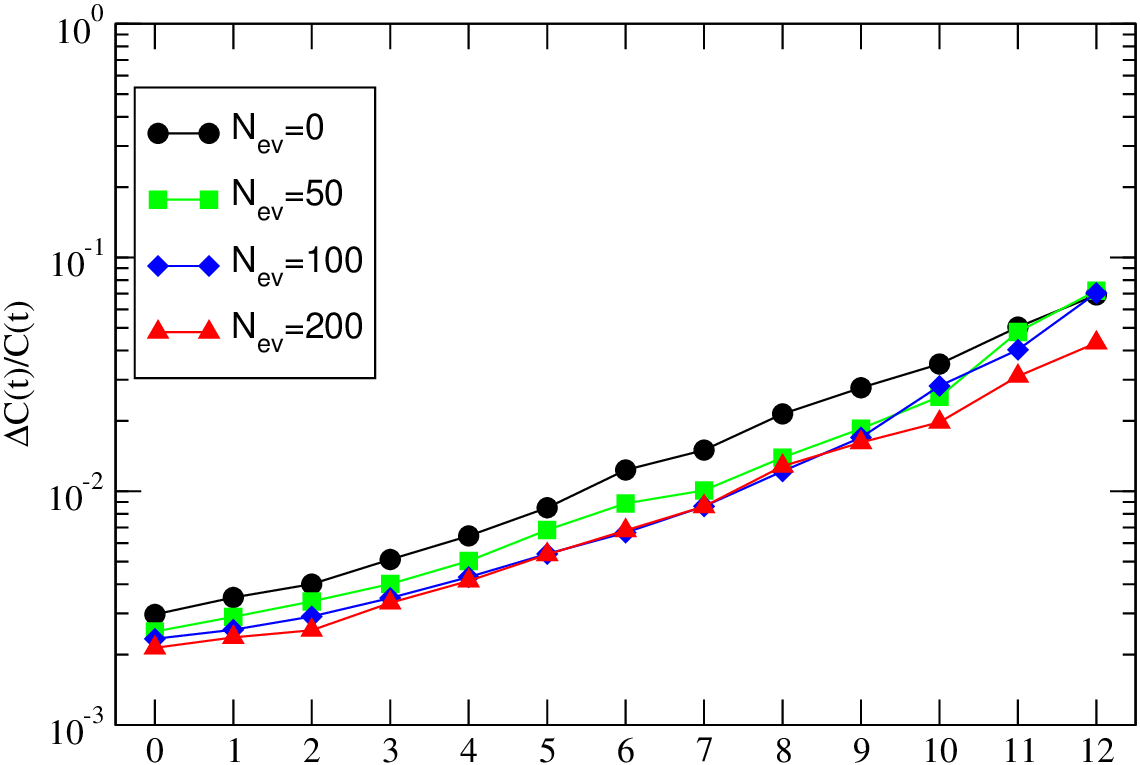}
\includegraphics[width=7.5cm,clip]{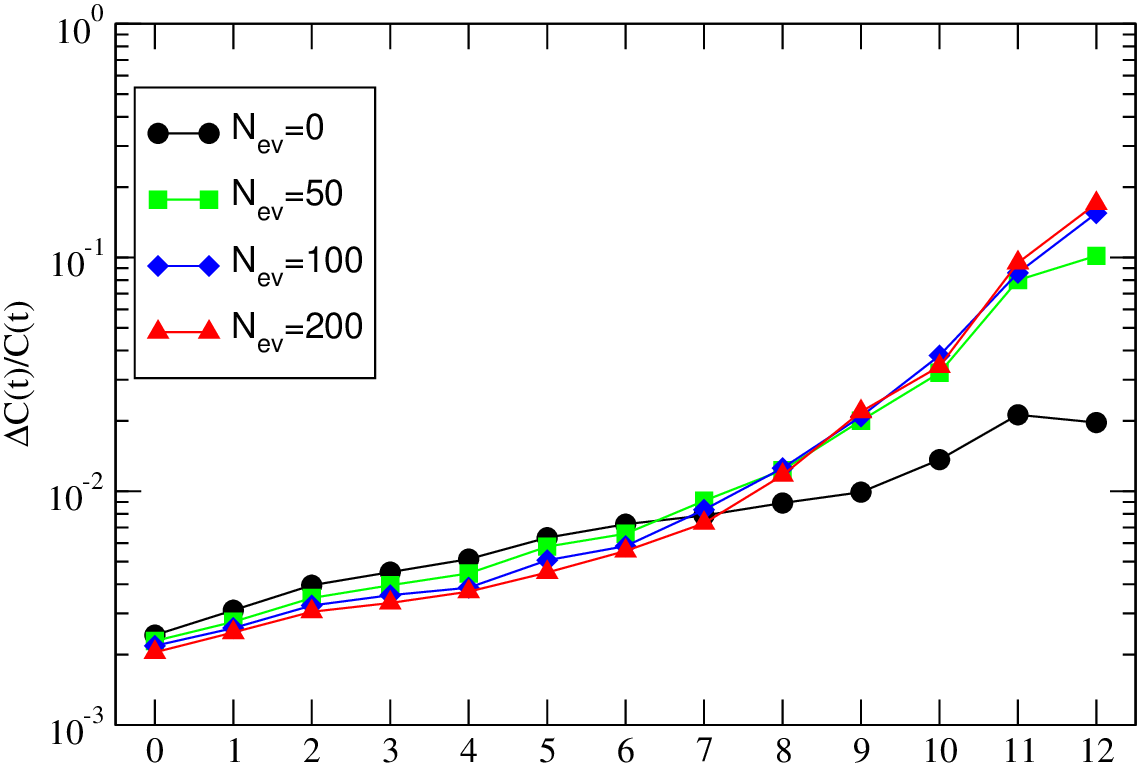}
\caption{
The time dependence of the noise to signal ratio for $\kappa=0.1464$
(left panel) and $\kappa=0.1430$ (right) at $\beta=1.80$ with 40
configurations. 
}
\label{fig:Nev}
\end{figure}

In Figure~\ref{fig:Nev} we show the comparison of the
noise to signal ratio of the 2-point functions with
different number of low eigenmodes, $N_{\rm ev}=0$, 50, 100, 200 for
$\kappa=0.1464$ and $\kappa=0.1430$ at $\beta=1.80$. 
For the smallest quark mass, the statistical error of the 2-point
function is 1.5--2 times improved as $N_{\rm ev}$ is changed from
0 to 200.
While this is not a drastic improvement, comparing the costs to
determine the low-lying eigenmodes and to solve the quark propagator
(the latter is 8--10 times larger than the former),
there is still an advantage to adopt the low-mode averaging.
Projecting out the low-lying modes also improves the cost of solving
the quark propagator.
These effects are amplified as going to smaller quark mass region.

In the region of larger light quark mass, however, the situation
is different.
The right panel of Fig.~\ref{fig:Nev} shows the noise to signal ratio
of the 2-point function for $\kappa=0.1430$. 
The noise to signal ratio for $t <7$ achieves about factor 1.3 improvement
in the statistical error as we change $N_{\rm ev}$ from 0 to 200.
For $t>7$, on the contrary, the noise to signal ratio with
$N_{\rm ev}\neq 0$ starts to grow more rapidly than that with
$N_{\rm ev}=0$ which keeps to grow steadily.
As the result, the low-mode averaging deteriorates the statistical
accuracy at this and larger light quark masses.

\subsection{High mode and low mode contributions to the 
noise}

\begin{figure}[tb]
\includegraphics[width=7.5cm,clip]{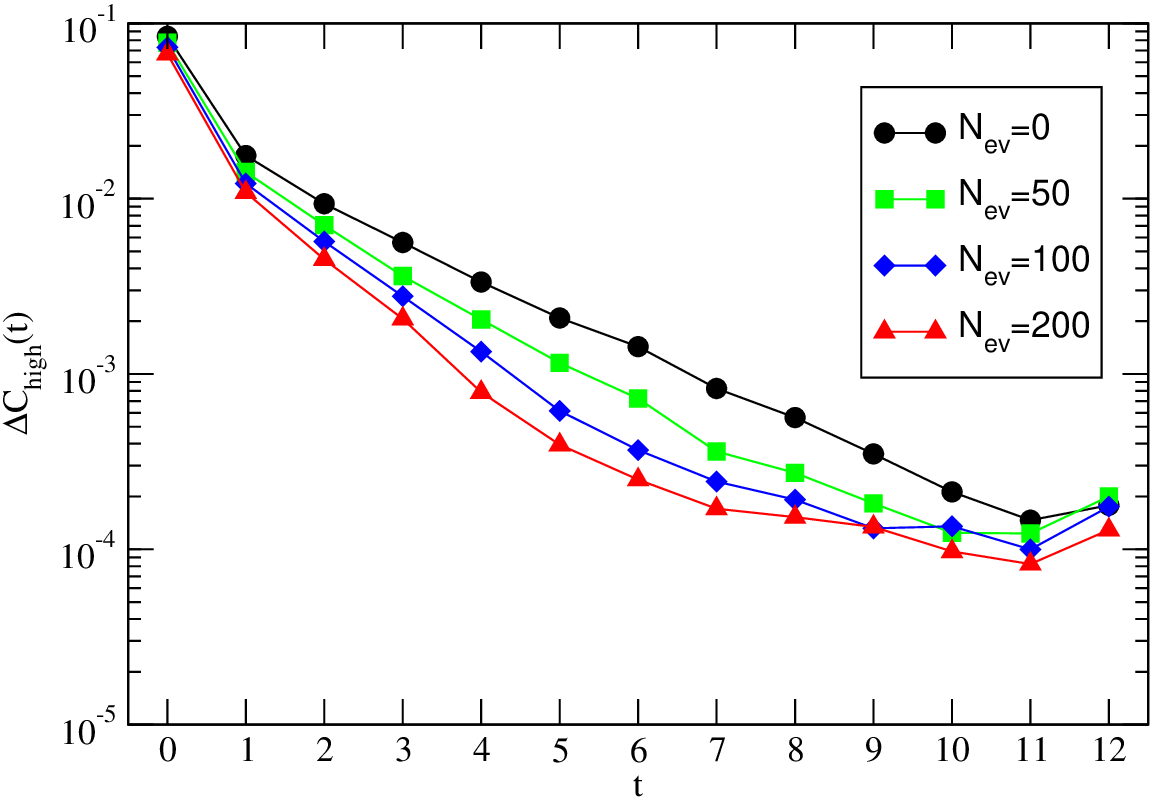}
\includegraphics[width=7.5cm,clip]{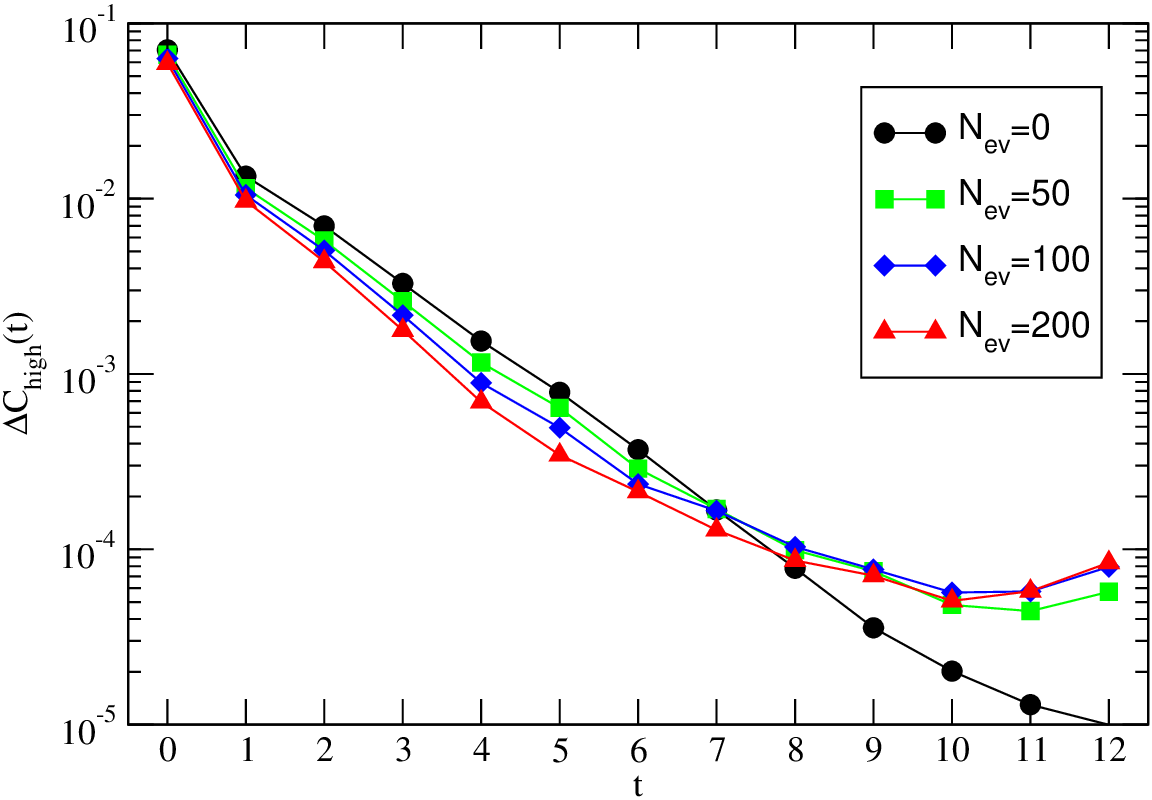}
\caption{
The time dependence of the noise from the high mode
part for $\kappa=0.1464$ (left panel) and $\kappa=0.1430$ (right)
at $\beta=1.80$ with 40 configurations. }
\label{fig:highmode_error_Nev}
\end{figure}

To investigate the origin of this behavior, 
we examine the high mode and low mode contributions 
to the error of correlator separately.

In general, by projecting out larger number of low modes, the high
mode part decreases. Therefore one might naively expect 
that the error also decreases. This is indeed the case for 
$\kappa=0.1464$.  However for $\kappa=0.1430$ such a naive expectation 
does not hold. Figs.~\ref{fig:highmode_error_Nev} shows the time dependence  
of the error for the high mode part of the 2 point correlator 
with various values of $N_{\rm ev}$ at $\kappa=0.1464$ and 
$\kappa=0.1430$.
As is displayed in the right panel of Fig.~\ref{fig:highmode_error_Nev},
although the error of the high mode 
contribution to the correlator at $\kappa=0.1430$ 
does decrease at $t<7$ with larger $N_{\rm ev}$, 
the errors of the high mode part at $t>7$ with $N_{\rm ev}=50,100,200$ exceed 
that with $N_{\rm ev}=0$. 
This clearly indicates that for $t>7$ both the high and low mode parts
of the correlator individually have large errors,
but when they are combined the error 
of the total correlator becomes small. 
In this situation if the low mode part is improved by the low mode averaging,
which reduces the error by certain factor,
the error of the high mode part of the same
size remains unreduced and dominates the error of the correlator. 

\begin{figure}[tb]
\centerline{
\includegraphics[width=7.5cm,clip]{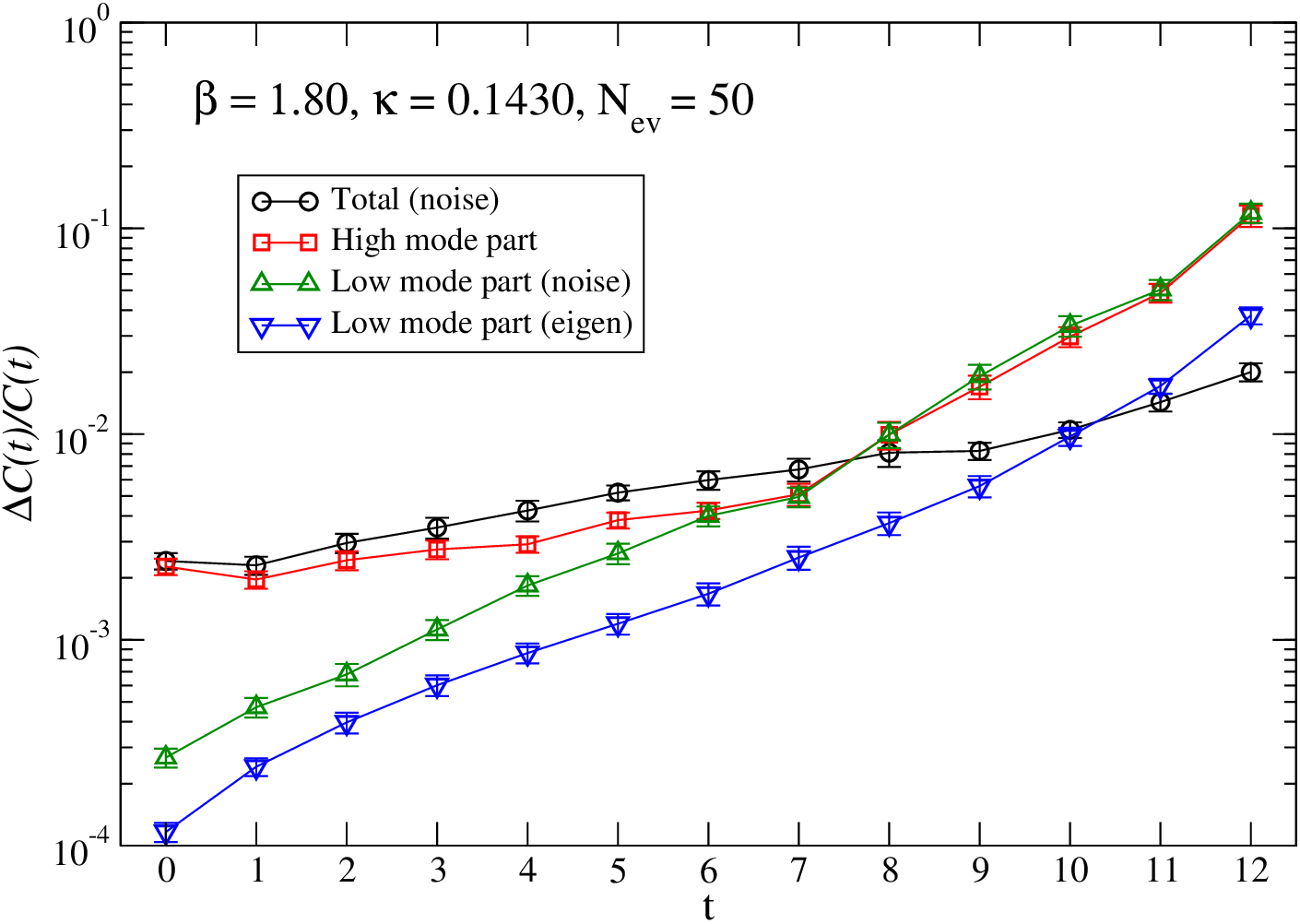}\\
}
\caption{
The noise to signal ratio of the low and high mode
parts for the 2-point correlators for $\kappa=0.1430$
at $\beta=1.80$ with 40 configurations.
The signal part is always taken to be the total 
correlator using only the noisy estimator.
The projection is made with 50 eigenmodes.
}
\label{fig:error_Nev1}
\end{figure}

\begin{figure}[tb]
\includegraphics[width=7.5cm,clip]{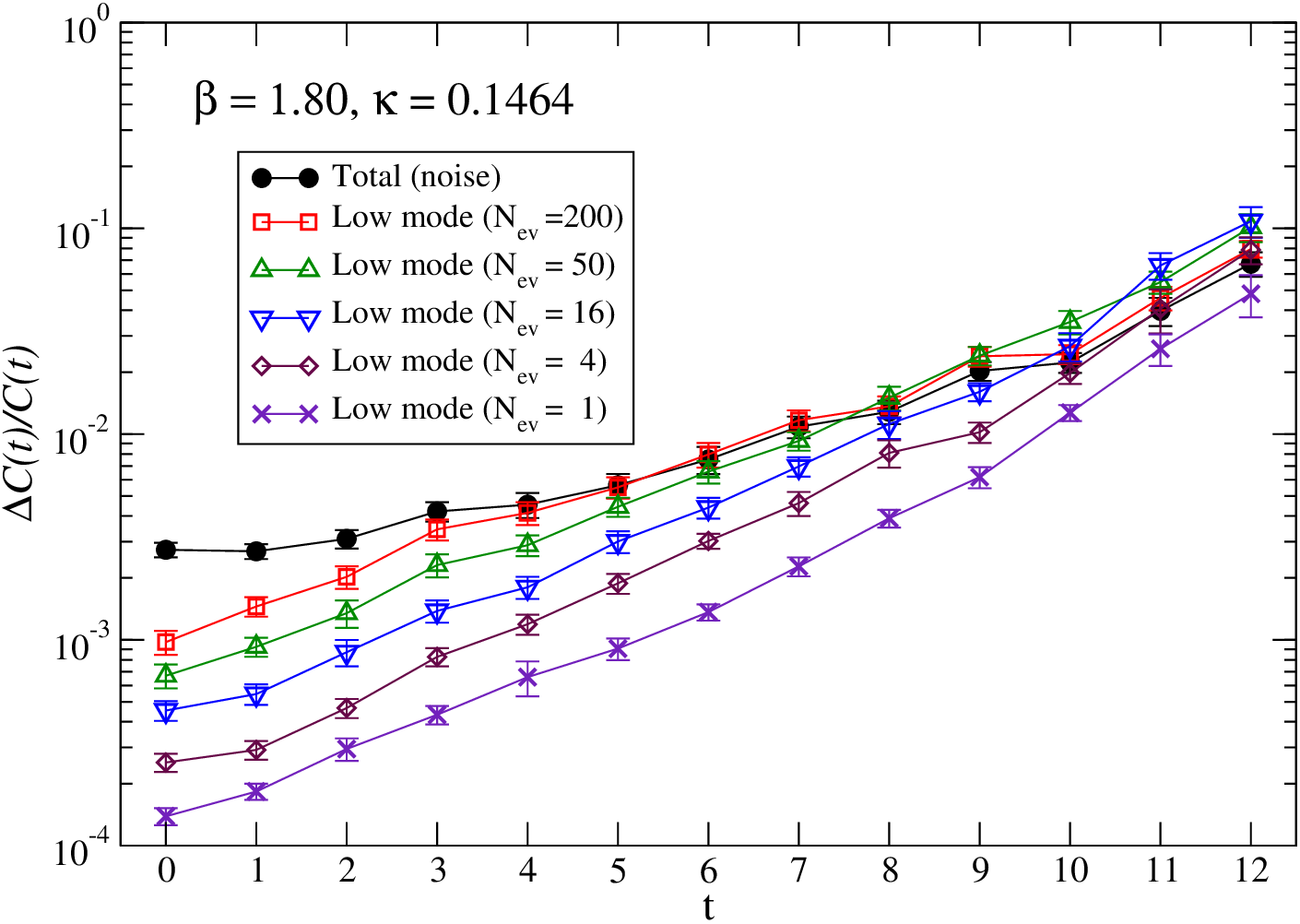}
\includegraphics[width=7.5cm,clip]{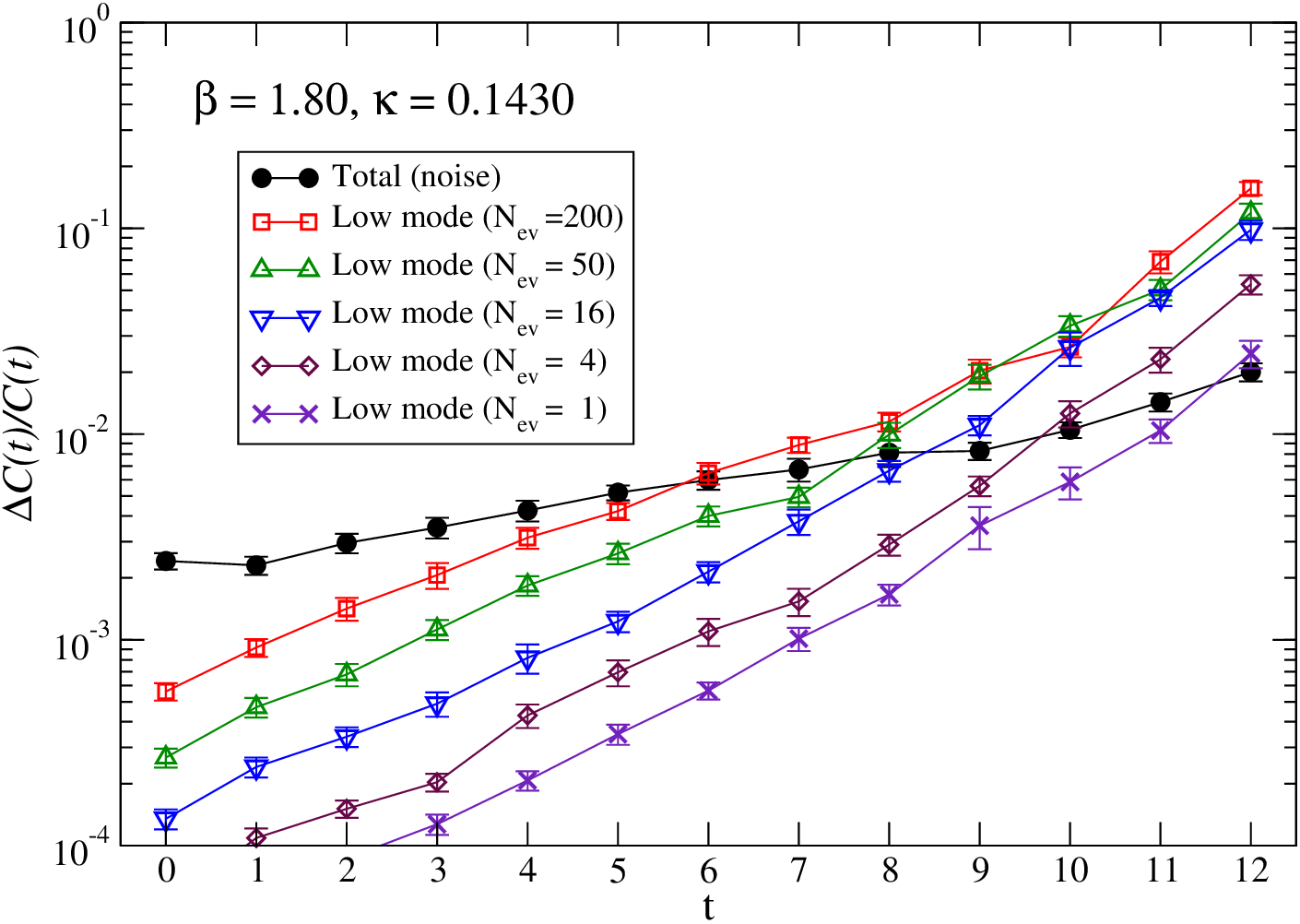}
\caption{
The noise to signal ratio of the low and high mode
parts for the 2-point correlators at $\beta=1.80$ with 40
configurations.
The signal part is always taken to be the total 
correlator using only the noisy estimator.
The left and right panels show the dependence of the error of
`low(noise)' on $N_{\rm ev}$ for $\kappa=0.1464$ and $\kappa=0.1430$,
respectively.}
\label{fig:error_Nev2}
\end{figure}

To see it more clearly, let us decompose the 2-point correlator 
computed by the noisy estimator (corresponding to $N_{\rm ev}$=0)
into the high and low mode parts.
The total correlators are computed only with the noisy 
estimator (denoted as `total(noise)').
The high and low mode parts (`high(noise)' and 'low(noise)') 
are separately computed using the exactly same random source
as for the total correlators but projected into 
the high and low mode spaces with the projection operators 
$P_1$ and $P_0$, respectively. 
Figure~\ref{fig:error_Nev1} displays the statistical errors from  
the low and high mode parts normalized with the total correlator,
in the case of $N_{\rm ev}=50$ at $\kappa=0.1430$.
For comparison we also show the error of the low mode part determined
with low mode averaging (`low(eigen)'). 
This figure confirms that the fluctuations of the low and high mode
parts are almost the same size and compensate in the total correlator
so as to give much smaller error.
The errors of the low mode parts,
$E_{noise}(t)$ and $E_{eigen}(t)$, exponentially grow with similar
rates, while different from that of the total correlator.

Such a behavior continues as we decrease $N_{\rm ev}$ even down to
a few $N_{\rm ev}$.
The right panel of Fig.~\ref{fig:error_Nev2} shows the case of
$N_{\rm ev}=1$, 4, 16, 50, 200 for $\kappa=0.1430$, where the fluctuations
of the low mode part, $E_{noise}(t)$, grows similar
rates, while absolute values are shifted downward.
The left panel of Fig.~\ref{fig:error_Nev2} shows the case 
for $\kappa=0.1464$.
We observe that the low mode part (`low(noise)') 
does not exceed the signal by large amount, which explains the 
behavior observed in Fig.~\ref{fig:Nev}.

The reason why projecting into the low or high mode part 
provides a drastic enhancement of the error is still unknown.
Although the phenomena themselves are quite interesting and deserves
for further studies, in this paper we restrict ourselves within their
implication to applicability of the all-to-all propagator technique to
the static heavy-light system.

\subsection{When is the low mode averaging efficient?}

We have seen that the low mode averaging
is efficient only if the error of the noisy
estimator (not the correlator itself) is dominated by the low mode part.
Once the errors from the high and low mode parts of the correlator 
start to exceed the error of the total correlator, 
the low mode averaging is no longer effective 
but it makes the situation even much worse. 
Our result implies that at $\beta=1.80$ and $\kappa<0.1430$,
the rapid growth of the error at $t\geq 7$ in Fig.~\ref{fig:Nev}
signals the breakdown of the above condition.
For $\kappa\leq 0.1430$, the noisy estimator without the low mode
averaging works better.
Thus the low mode averaging is only efficient in the small
quark mass region.
However, since we have already taken the data and they provided
satisfactory statistical accuracy of 2\% level,
we adopted the result with the low mode averaging propagator
at $\beta=1.80$.
As for $\beta=1.95$, the low mode averaging
has not provided sufficient statistical accuracy for $\kappa=0.1375$.
Thus we adopted the noisy estimator without the low mode averaging
at this $\kappa$ as was already noted in the previous section.

\begin{figure}[tb]
\begin{center}
\includegraphics[width=8cm,clip]{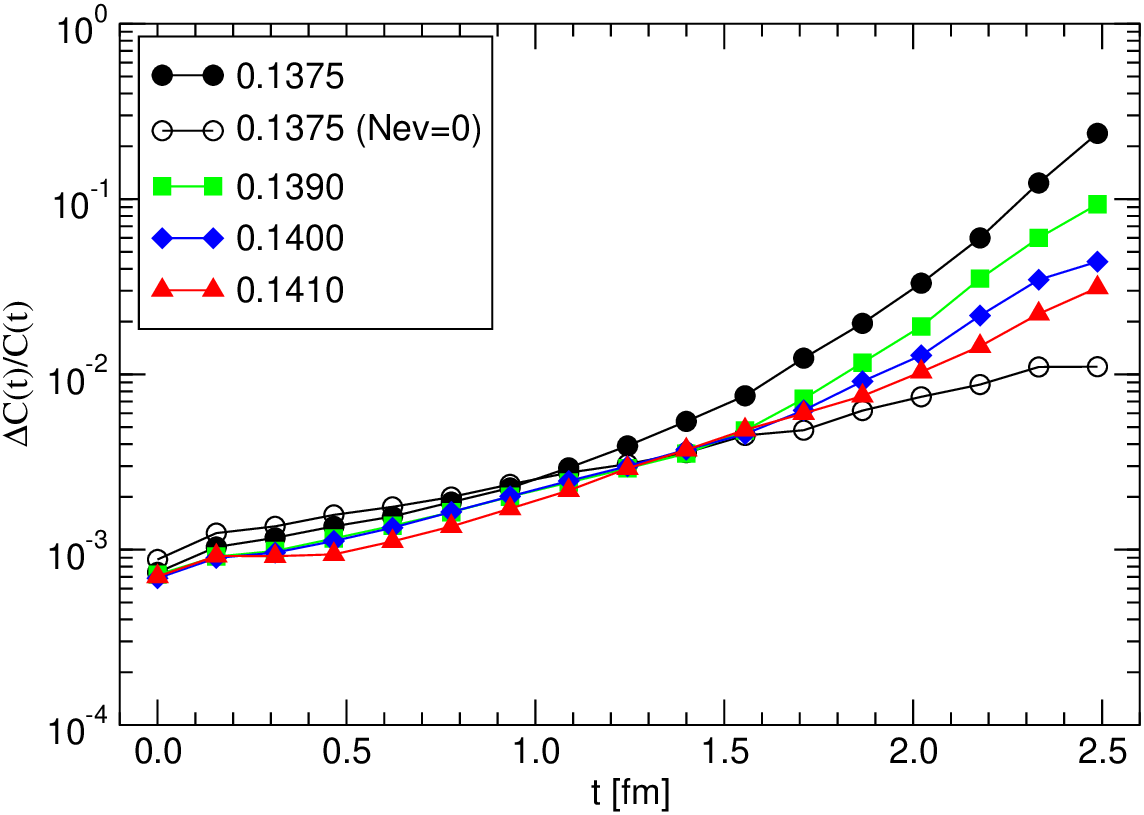}
\caption{
The time dependence of the noise to signal ratio of the 2-point
correlators at $\beta=1.95$.
The results are determined with $N_{\rm ev}=200$, while
for $\kappa=0.1375$ the $N_{\rm ev}=0$ result is also displayed.}
\label{fig:noise-to-signal_atbeta1.95}
\end{center}
\end{figure}

The results at $\beta=1.95$ are displayed in
Figure~\ref{fig:noise-to-signal_atbeta1.95}.
The figure shows the noise to signal ratio of the 2-point correlators
at each $\kappa$ against $t$ in physical units.
For $\kappa=0.1375$ both the results with $N_{\rm ev}=0$ and 200
are shown, and the former indeed exhibits smaller statistical
error.
For all the values of $\kappa$ with $N_{\rm ev}=200$, the slopes of
the exponential growth rate of the noise to signal  ratio
change around $t\sim 16$, and beyond that $t$ the slopes become
steeper as the quark mass increases.
This behavior is clearly explained with the breakdown mechanism of
the low mode averaging mentioned above.

We can also extract a hint on the lattice spacing dependence of the
statistical accuracy by comparing $\beta=1.80$ and $\beta=1.95$.
Comparison of Figs.~\ref{fig:Nev} and 
\ref{fig:noise-to-signal_atbeta1.95} implies that the noise to signal
ratio is similar or even smaller for finer lattices.
This is partly explained by the fact that as going the finer lattices
one has the larger number of lattice points (if the volume is kept
unchanged) which are used for the all-to-all propagator.
As observed in Figs.~\ref{fig:meff_2pt_b1.80} and
\ref{fig:Nev}, the statistical errors at $\beta=1.80$ rapidly
increases beyond $t\sim 8$, which corresponds to $t\sim 1.6$ fm
in physical units.
Thus the low mode averaging breaks down almost at the same physical
distances at these two lattice spacings.
This implies that also on finer lattices of $a\sim 0.1$ fm,
one can extract precise values of $B^*B\pi$ coupling from the
region $t< 1.6$ fm by applying the same methods as this work,
while careful tuning of the smearing function would be indispensable.

\section{Conclusion}
\label{Conclusion}

\begin{figure}[tb]
\begin{center}
\includegraphics[width=9cm,clip]{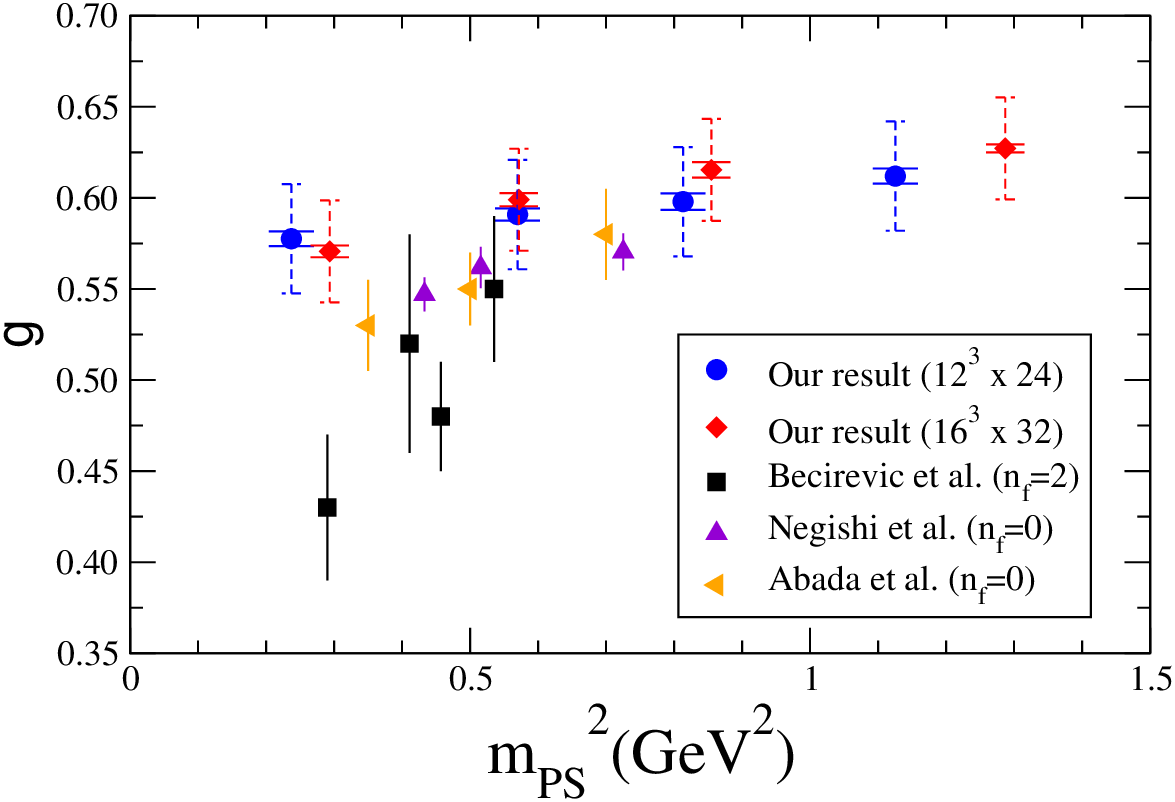}
\caption{
Comparison of $\hat{g}_{\infty}$ with other
calculations~\cite{Abada:2003un,Becirevic:2005zu,Negishi:2006sc}.
In our results, the small and large errors represent the statistical
error and the perturbative error, respectively. }
\label{fig:comparison}
\end{center}
\end{figure}

In this paper, we computed the $B^* B\pi$ coupling on unquenched lattices 
using the HYP smearing and the all-to-all propagators.
Using the low mode averaging with 200 eigenmodes, the statistical
errors are kept sufficiently small for smaller quark masses.
On the other hand, as was investigated
in Sec.~\ref{sec:low_mode_averaging} in detail, the low mode averaging
is not efficient for larger light quark mass region, where the simple
noisy estimator provides better precision.
In either case, the statistical error is controlled below
2\% level in the chiral limit.
We obtained consistent results at two lattice spacings.
Our best estimate of the $B^*B\pi$ coupling in the static limit is 
represented in Eq.~(\ref{eq:final_result}).
Figure~\ref{fig:comparison} compares our results with other recent
works on the $B^*B\pi$ coupling
\cite{deDivitiis:1998kj,Becirevic:2005zu,Negishi:2006sc}.
The improvement in statistical precision is drastic, which proves
the power of the improvement techniques employed in this paper.

For future prospects, better control over the systematic error from
the chiral extrapolation is indispensable.
For this purpose, the configurations with dynamical overlap fermions
by JLQCD collaboration would be a good choice
\cite{Matsufuru:2007uc,Hashimoto:2007vv,Kaneko:2006pa}.
In order to obtain $\hat{g}_b$ at the physical bottom quark mass, one
needs to understand the mass dependence of $\hat{g}$.
Simulations with the charm quark mass region and interpolation with
the static limit are desired.
The methods adopted in this work are in principle also applicable to
other weak matrix elements of the $B$ mesons, such as $f_B$, $B_B$,
and the form factors, and expected to provide high precision
results required in precision flavor physics.

\section*{Acknowledgments}
We would like to thank S.~Aoki, M.~Della~Morte, N.~Ishizuka, 
C.~Sachrajda, T.~Umeda for fruitful discussions.
We are also grateful to S.~Fajfer and J.~Kamenik for useful comments.
We acknowledge JLDG for providing with unquenched
configurations from CP-PACS collaboration.  
The numerical calculations were carried out on the vector
supercomputer NEC SX-8 at Yukawa Institute for Theoretical Physics, 
Kyoto University ,Research Center for Nuclear Physics, 
Osaka University and also Blue Gene/L at
High Energy Accelerator Organization (KEK).
The simulation also owes to a gigabit network SINET3 
supported by National Institute of Informatics,  
for efficient data transfer supported by JLDG.
This work is supported in part by the Grant-in-Aid of the
Ministry of Education (Nos. 19540286, 19740160).

\end{document}